\documentclass{egpubl}
\usepackage{eurovis2025}

\SpecialIssuePaper         

\CGFccby

\usepackage[T1]{fontenc}
\usepackage{dfadobe}

\usepackage{cite}  
\BibtexOrBiblatex
\electronicVersion
\PrintedOrElectronic
\ifpdf \usepackage[pdftex]{graphicx} \pdfcompresslevel=9
\else \usepackage[dvips]{graphicx} \fi

\usepackage{egweblnk}

\usepackage{pdfpages}

\title[InterChat]{InterChat: Enhancing Generative Visual Analytics using Multimodal Interactions}

\author[J. Chen, J. Wu et al.]{
\parbox{\textwidth}{\centering
  Juntong Chen\thanks{
     Juntong Chen and Jiang Wu conducted this research as visiting scholars in Dongyu Liu's group at the University of California, Davis.
  }$^{1}$\orcid{0000-0001-9343-4032}, 
  Jiang Wu$^{\dagger 1}$\orcid{0009-0001-8831-1473}, 
  Jiajing Guo $^{23}$\orcid{0000-0003-0511-136X}, 
  Vikram Mohanty $^{23}$\orcid{0000-0001-6296-3134}, 
  Xueming Li $^{4}$\orcid{0009-0006-2899-2832},\\ 
  Jorge Piazentin Ono $^{23}$\orcid{0000-0002-2424-0186}, 
  Wenbin He $^{23}$\orcid{0000-0002-5376-5803}, 
  Liu Ren $^{23}$\orcid{0009-0002-1813-8844}, 
  Dongyu Liu$^{1}$\orcid{0000-0002-8915-2785} 
}
\\
{\parbox{\textwidth}{\centering %
  $^1$University of California, Davis, USA\\
  $^2$Bosch Research North America, USA\\
  $^3$Bosch Center for Artificial Intelligence (BCAI)\\
  $^4$Robert Bosch GmbH\\
}
}
}

\usepackage{enumitem}

\definecolor{dypink}{HTML}{ec008c}

\newcommand{\bf}[1]{\textbf{#1}}

\usepackage{marginnote}
\usepackage{adjustbox}

\global\marginparsep=12pt
\newcommand{\sidecomment}[1]{%
  \ifdefined\revise%
  \marginnote{%
    \textcolor{Magenta}{%
      \adjustbox{minipage=0.25\marginparwidth,fbox}{%
          \scriptsize#1%
      }%
    }%
  }%
  \fi%
}%

\newcommand{\revision}[1]{%
  \ifdefined\revise%
  {\hypersetup{allcolors=Magenta}\color{Magenta}#1}%
  \else%
  #1%
  \fi%
}%

\newcommand{\revisiontext}[1]{%
  \ifdefined\revise%
  {\textcolor{Magenta}{#1}}%
  \else%
  #1%
  \fi%
}%

\newcommand{\revisionbox}[1]{%
  \ifdefined\revise
  \setlength{\fboxrule}{0.5pt} 
  \setlength{\fboxsep}{1pt} 
  \fcolorbox{Magenta}{white}{#1}
  \else
  #1
  \fi
}

\begin{document}


\maketitle

\begin{abstract}

The rise of Large Language Models (LLMs) and generative visual analytics systems has transformed data-driven insights, yet significant challenges persist in accurately interpreting users’ analytical and interaction intents. While language inputs offer flexibility, they often lack precision, making the expression of complex intents inefficient, error-prone, and time-intensive.
To address these limitations, we investigate the design space of multimodal interactions for generative visual analytics through a literature review and pilot brainstorming sessions. Building on these insights, we introduce a highly extensible workflow that integrates multiple LLM agents for intent inference and visualization generation.
We develop InterChat, a generative visual analytics system that combines direct manipulation of visual elements with natural language inputs.
This integration enables precise intent communication and supports progressive, visually driven exploratory data analyses.
By employing effective prompt engineering, and contextual interaction linking, alongside intuitive visualization and interaction designs, InterChat bridges the gap between user interactions and LLM-driven visualizations, enhancing both interpretability and usability.
Extensive evaluations, including two usage scenarios, a user study, and expert feedback, demonstrate the effectiveness of InterChat. Results show significant improvements in the accuracy and efficiency of handling complex visual analytics tasks, highlighting the potential of multimodal interactions to redefine user engagement and analytical depth in generative visual analytics.

\ccsdesc[500]{Human-centered computing~Interactive systems and tools}
\ccsdesc[400]{Human-centered computing~Visual analytics}
\ccsdesc[300]{Computing methodologies~Natural language processing}

\printccsdesc
\end{abstract}

\section{Introduction}

Recent advancements in large language models and generative models have fueled significant interest in their application to visual analytics, a field we term \textbf{Generative Visual Analytics}.
This paradigm empowers users to articulate analytical needs through natural language, facilitating intelligent visual-driven data exploration, interpretation, and summarization\cite{yang2024foundation,ye2024generative,zhao2024leva,shen2023towards}.
By enabling users to specify diverse queries without requiring specialized technical expertise, generative visual analytics enhances both accessibility and flexibility.
However, relying solely on natural language inputs presents challenges in precisely conveying complex and evolving analytical intents. The nuanced and ambiguous nature of language can lead to inefficiencies, inaccuracies, and increased effort, particularly in iterative, real-world analytics workflows demanding precise communication.

Visual analytics frequently require users to identify patterns within visualizations and conduct subsequent analyses based on those observations. Language-only interactions often fall short in precisely conveying such patterns. For instance, describing a fluctuation in a time-series chart or identifying clusters in a scatterplot can be cumbersome and prone to errors. Similarly, performing complex analytical tasks---such as comparing fluctuations in time-series data across multiple periods---demands detailed prompts that are not only time-consuming to construct but also susceptible to inaccuracies.
Traditional visual analytics systems rely on predefined interaction workflows to simplify specific tasks but lack flexibility for a broader range of analytical needs. While LLMs offer flexibility in articulating diverse analytical intents, they confine users to language-based interactions, which can become a bottleneck for complex tasks.
For example, replicating the aforementioned time-series comparison might necessitate a detailed prompt like:
``\textit{Compare the fluctuations in data from {\small\texttt{<start\_time\_a>}} to {\small\texttt{<end\_time\_a>}} with those from {\small\texttt{<start\_time\_b>}} to {\small\texttt{<end\_time\_b>}} using {\small\texttt{<chart\_type>}}}.'' This approach, though flexible, is less intuitive and efficient compared to direct, visually driven interactions (e.g., drawing selection boxes).

To overcome these challenges, we introduce \textbf{InterChat}, a generative visual analytics system that leverages multimodal interactions to bridge users and LLMs.
Multimodal interaction, a concept rooted in Human-Computer Interaction, combines different input modalities, such as natural language, gestures, and direct manipulation of visual elements, to enhance the naturalness and flexibility of user interactions~\cite{turk2014multimodal}.
In InterChat, we specifically integrate traditional visualization interactions on a 2D interface (e.g., clicking, dragging, and selecting data points) with natural language inputs.

\revision{\sidecomment{R3.4}%
The design of InterChat is informed by pilot brainstorming sessions with eight VIS/HCI researchers and existing literature, resulting in a design space comprising three key components: interaction, intent, and instruction.
}%
Building on this foundation, InterChat integrates multimodal inputs via a multi-agent architecture, enabling users to articulate complex analytical intents with precision. The system incorporates rich interactivity into generated visualizations, allowing for more efficient, visually driven analyses.

The evaluation comprises three parts. First, we explored two real-world usage scenarios using Netflix stock prices and steel manufacturing data. Second, we conducted a user study with ten participants, gathering both qualitative and quantitative feedback on task completion time, visualization accuracy, intent inference, and answer correctness. Third, we obtained additional feedback from manufacturing industry experts. The results show that multimodal interactions significantly improve accuracy and efficiency, especially for complex tasks requiring multiple interactions.

In summary, our contributions are as follows:

\begin{itemize}
    \item The design space for applying multimodal interactions in generative visual analytics, consisting of intent space, interaction space, \revision{and instruction space.%
    }%
    \item An extensible workflow integrating multi-agent LLM architecture to enhance intent inference and visualization generation.

    \item A generative visual analytics system InterChat that combines direct manipulation of visual elements with natural language inputs, enabling precise conveying of complex analytical intents.

    \item A comprehensive evaluation conducted through two real-world usage scenarios, a rigorous user study, and expert feedback.
   
\end{itemize}
\vspace{-0.75em}
\section{Related Work}

\subsection{Multimodal Interactions}

\noindent
Systems with multimodal interactions can enhance the naturalness and flexibility of human-computer interaction by allowing simultaneous or sequential use of multiple input modes \cite{turk2014multimodal,norris2004analyzing}.
For example, those in Extended Reality (XR) environments usually support combining gestures with voice commands \cite{lee2008wizard,yang2022hybridtrak,zimmerer2020finally}, exemplified by ``Put-that-there'' \cite{bolt1980put}, where gestures select objects and voice commands execute actions.
Similarly, eye tracking can also be combined with speech input in XR environments \cite{alhargan2017multimodal,brone2023mobile}.
Mobile devices are another ideal application scenario for multimodal interactions \cite{williamson2007shoogle,brewster2003multimodal,yang2024reactgenie}.
``Data@hand'' \cite{kim2021data}, a typical example combining touch interactions on mobile devices and speech input, enables better interpretation of user commands.

A notable pattern in existing systems is the prevalent adoption of a ``language input + X'' approach, favored due to the simplicity and scalability of language input. The additional modality---whether gesture, touch, or another form---typically refines language commands or simplifies their expression.

Our study studied how to integrate the ``language input + X'' paradigm into the generative visual analytics process and make the overall workflow intuitive, efficient, and precise.
\subsection{Visual Analytics Using Natural Language}

Prior to the advent of LLMs, researchers applied natural language processing (NLP) techniques in visual analytics and developed visualization-oriented natural language interfaces (V-NLI) \cite{kavaz2023chatbot}.
A recent survey by Shen et al. \cite{shen2023towards} summarizes the state-of-the-art techniques used for each step in the process of augmenting visual analytics with natural language.
DataBreeze \cite{Srinivasan2021interweaving} allows users to speak commands for selected visual units, such as sorting and filtering.
NL4DV \cite{Narechania2021nl4dv} generates several candidate Vega-Lite specifications based on a tabular dataset and a query in the natural language given by users.
Wang et al. \cite{wang2023towards} introduce authoring-oriented NLIs to facilitate visualization creation.
SlopeSeeker \cite{bendeck2024slopeseeker} enables natural language querying and analysis of time series trends by mapping quantifiable data patterns to meaningful trend descriptors.

However, due to the limited capabilities of earlier NLP technologies and the inherent ambiguity of natural language, most existing works adopt template-based or rule-based methods to process user commands.
Consequently, users' expressions of intent are often confined to the command space supported by the system, hindering the full utilization of natural language’s advantages.
To overcome this limitation, we investigated the design spaces and methods for leveraging multimodal interactions to enhance the interactivity and effectiveness of generative visual analytics systems.

\subsection{LLMs for Visualization}

\noindent
Although the use of natural language for visualization has been developed for decades \cite{shen2023towards,wu2022ai}, innovations in LLMs bring new opportunities and challenges \cite{vazquez2024are,ye2024generative,zengAdvancing2025}, particularly in visualization generation and interpretation.

LLMs enable the creation of visualizations from user prompts.
Tools such as LIDA \cite{dibia2023lida}, ChartGPT \cite{tian2024chartgpt}, LLM4Vis \cite{wang2023llm4vis}, and Prompt4Vis \cite{li2024prompt4vis} automate insight extraction and optimize natural language to chart conversion, leveraging techniques such as reasoning processes and multi-objective example mining.
NL2Color \cite{shi2024nl2color} refines chart palettes based on user input.
ChartSpark \cite{Xiao2024284} enhances visualization quality by interpreting semantic context.
Shen et al. \cite{shen2024from} and Ying et al. \cite{ying2024reviving} employ LLMs to create animated data charts.

LLMs enhance visualization interpretation through various applications, such as generating captions \cite{liew2022using,tang2023vistext,Ko2024natural}, synchronizing narratives with animations \cite{shen2024data}, creating textual narratives for data-driven articles \cite{sultanum2023datatales}, supporting multi-stage exploration and summarization \cite{zhao2024leva}, and training specialized models for various tasks \cite{masry2023unichart,han2023chartllama}.

Our research distinguishes itself by integrating multimodal interactions into LLM-driven visual analytics. By enriching generated visualization with dynamic interactions and supporting a progressive analysis workflow, we introduce a new approach to generative visual analytics that enhances both interpretability and usability.

\vspace{-0.5em}
\section{Informing the Design}

\subsection{Pilot Brainstorming Session}

\revision{

\sidecomment{R1.2 \ R3.3 \ SR5}

Our system is intended for users with intermediate to advanced data analysis expertise who are familiar with exploratory analysis and frequently use charts for analytical tasks.
Since no prior work explores LLM-empowered multimodal interactions for visual analytics, we conducted a pilot brainstorming session to gather use cases and insights to guide our system design.

\sidecomment{R1.1 \ R2.1 \ R3.2 \ R4.7}
We recruited eight participants (4 males, 4 females) with at least three years of visual analytics experience. Each had developed a VA system for various domains, while only two had prior experience with multimodal interactions in VA. Additionally, all participants used LLM-based conversational tools at least once a week, incorporating them into their workflow for tasks such as text writing (7/8), code generation (5/8), and data analysis (3/8).

The session started with introductions about multimodal interaction, followed by a demographic study. Then, participants were asked to brainstorm potential use cases for multimodal interactions in visual analytics, focusing on common scenarios involving 2D visualizations. The discussion was entirely conversational, without involving any specific dataset or system.
For each case, participants described: (1) an existing common chart type for this analysis scenario; (2) the analytical intent; (3) the required multimodal interactions (\textbf{MI}), combining direct chart manipulation with natural language (\textbf{NL}) input to an LLM like ChatGPT to achieve the intent, and (4) the use of NL input alone for the same goal. They then rated their preference between MI and NL-only options on a 7-point Likert scale (1 = MI is much worse; 7 = MI is much better). Finally, they shared their thoughts on generative visual analytics, MI’s benefits, and their expectations for a generative VA system.

We collected a total of 75 use cases, with an average MI preference score of 5.96 ($\textrm{SD} = 1.02$).
(1) Most participants rated MI as slightly better (5 points, 13/75), better (6 points, 29/75), or much better (7 points, 26/75) than NL.
(2) A few cases showed no preference 4 points, 5/75) or a slight preference for NL(3 points, 2/75). These instances involved more complex chart interactions, where participants expressed concerns about the LLM's ability of intent interpretation.
(3) No participant rated MI as worse (2 points) or much worse (1 point). However, as all cases were intentionally designed to utilize multimodal interactions, the results may be biased.

The collected use cases and participants' shared open insights on generative visual analytics have provided an initial foundation for defining our design space and requirements for InterChat, as detailed in the following sections.

}

\vspace{-0.3em}
\subsection{Design Space}

\begin{figure}[t]
  \centering
  \revisionbox{\sidecomment{R3.4}
    \includegraphics[width=\linewidth]{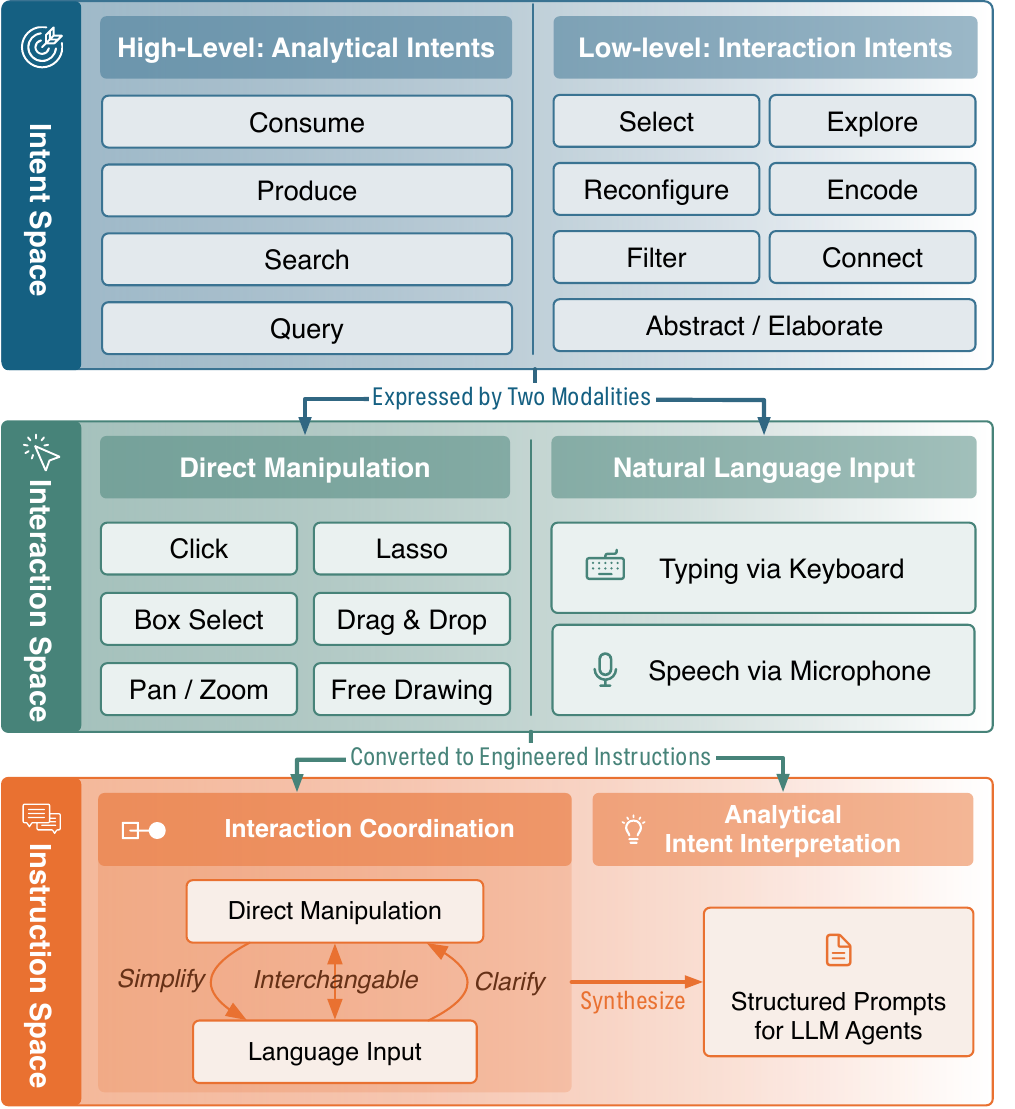}
  }
  \caption{%
    The design space of InterChat. The intent space defines users' goals within the generative VA system, with low-level interaction intents expressed via two modalities in the interaction space: Direct Manipulation and Natural Language Input.
    \revisiontext{%
      User's interaction, either through direct manipulation or natural language input (Interaction Space), are synthesized into structured prompts for LLM agents (Instruction Space, detailed in Section~\ref{sec:prompt_engineering}), enabling interaction coordination and precise analytical intent expression.
    }}

    \vspace{-0.7cm}
  \label{fig:design-space}
\end{figure}

\revision{
We structure the design space for multimodal interactions in generative visual analytics across three dimensions (Figure \ref{fig:design-space}): intent, interaction, and instruction.
}

\subsubsection{Intent Space}
\revision{The intent space is derived from the analytical intents identified in the collected cases and a literature review on typical visual analytics tasks\cite{yi2007toward,brehmer2013multi,edge2018beyond}.
This space reflects what users expect the VA system to accomplish. We categorize these expectations into high-level analytical goals (analytical intents) and the specific actions they aim to perform (interaction intents).}

\textbf{Analytical Intents} represent users' high-level objectives in data analysis.
According to Brehmer and Munzner’s taxonomy\cite{brehmer2013multi}, these include:
\textit{Consume} (e.g., \textit{present}, \textit{discover}, and \textit{enjoy}), \textit{Produce}, \textit{Search} (e.g., \textit{lookup}, \textit{browse}, \textit{locate}, and \textit{explore}), and \textit{Query} (e.g., \textit{identify}, \textit{compare}, \textit{summarize}).
Traditional visual systems support these intents through specific predefined interactions, but this can lead to increased development and learning costs as the number of supported intents grows. Visual analytics systems with natural language interactions allow users to express analytical intent language inputs, leveraging NLP to interpret requests, such as ``compare the highest values in these two datasets.''

\textbf{Interaction Intents} reflect users' low-level actions that manipulate visual elements.
Based on existing research\cite{yi2007toward}, these can be categorized into seven primary types: \textit{select}, \textit{explore}, \textit{reconfigure}, \textit{encode}, \textit{abstract/elaborate}, \textit{filter}, and \textit{connect}.
In a multimodal system, most interaction intents can be expressed through either natural language or direct manipulation, allowing flexibility. For instance, in a scatterplot, the \textit{select} intent could be fulfilled by saying “select all red points” or by lasso-selecting certain points.

A multimodal visual analytics system allows users to achieve multiple analytical and interaction intents simultaneously. For example, a user might lasso-select a cluster of points (\textit{select}) and then ask the system to ``make them red-highlighted'' (\textit{encode}), or select two clusters and request to ``compare them'' (\textit{compare}).

\subsubsection{Interaction Space}
The interaction space encompasses two modalities that users can employ to interact with visual analytics systems: direct manipulation (DM) and natural language (NL) input.

\revision{
\textbf{Direct Manipulation} \cite{edwin1985direct} is widely adopted by traditional visual analytics systems.
Among the 75 cases, participants used various direct manipulation techniques: \textbf{Click} (27/75) to \textit{select} visual elements before issuing an NL command, \textbf{Lasso} (19/75) to \textit{select} multiple small elements within a defined region, \textbf{Box Select} (15/75) to \textit{select} a specific range along the x- or y-axis, \textbf{Drag and Drop} (7/75) to \textit{connect} elements or \textit{reconfigure} charts, and \textbf{Pan/Zoom} (5/75) for \textit{filtering} and exploring data subsets. Additionally, in two cases, participants mentioned \textbf{Free Drawing} as a flexible way to express intent, such as sketching a trend on a line chart to identify similar patterns.
}

\textbf{Natural Language Input} allows users to interact with the visual analytics system using natural language.
Common interactions for language input include typing via keyboard or speech via microphone. This modality is discussed in detail in the language space.

\subsubsection{\revisiontext{Instruction Space}}
\sidecomment{R3.4}
\revision{
The instruction space highlights the LLMs' roles in multimodal interactions, including coordinating interactions and interpreting analytical intents. This space bridges user actions and model responses, enabling precise interpretation and execution of analytical intents.
}

\textbf{Interaction Coordination.}\revision{
LLM agents can coordinate language input with direct manipulations.
}

\textit{1) Language input can \textbf{clarify} direct manipulations}.
Direct manipulations come with ambiguous interaction intents and unclear analytical intents.
For instance, when a user lasso-selects some points in a scatter plot, the interaction intent could be \textit{select} or \textit{filter}.
Meanwhile, users may want to \textit{explore} other data fields of the selected instances.
Language input helps users clarify their intents.

\textit{2) Direct manipulations can \textbf{simplify} language input}.
Language input may fall short when expressing visual patterns.
For instance, users can select a subset of points in a scatter plot and refer to them using a phrase like ``these points'', without naming each point in natural language.
Users can also use a phrase like ``this trend'' to refer to a trend drawn freely, without describing the trend in detail.

\textit{3) Language input and direct manipulations are \textbf{interchangable}}.
Some intents can be expressed through either NL or DM. For example, in a scatterplot where  outliers are highlighted in red, selecting outliers can be done by lasso-selecting them or by stating ``select all red points.''

Such flexibility of interactions leads users to focus on the analysis rather than how to use the system.

\revision{
\textbf{Analytical Intent Interpretation.}
\sidecomment{R3.4}
A generative VA system with multimodal interactions should allow users to express analytical intents using natural language and direct manipulations simultaneously. To enable LLMs to handle these modalities, we use a multi-agent LLM architecture, where we use two agents for manipulation descriptor generation and trigger phrase extraction (Section \ref{sec:intent_guessing}) and a third agent for visualization generation (Section \ref{sec:prompt_engineering}). The visualization generation agent takes structured prompts from the first two agents, along with the user’s NL command and the current visualization context, to produce the needed output.
The structured prompt expression of users' interactions reveals the relationships between each direct manipulation and natural language input, clearly conveying users' analytical intents.
}

\subsection{Requirement Analysis}
\revision{\sidecomment{R3.2}%
During the brainstorming session, participants expressed challenges and expectations regarding multimodal interactions in generative visual analytics. A key concern was correctness, both in terms of the generated visualization code and the interpretation of analytical intents. When asked about the participants' reason for preferring NL to MI, they expressed concerns about the LLM's ability to accurately extract selected data instances or numerical values. Many also emphasized the need for generalizability, suggesting the system should adapt to diverse datasets and tasks.
Taking all these into consideration, we identified the following design requirements for an effective generative VA system with multimodal interactions.
}%

\begin{enumerate}[label={\bf R{{\arabic*}}}]
    \item \label{req:generalizability}
    \textbf{Generalize to Different Datasets and Domains.}
    The system should support generalizable analysis capabilities, enabling users to analyze their own datasets across various domains and discover data insights.

    \item \label{req:interactivity}
    \textbf{Generate Visualizations with Intuitive Interactions.}
    The system should produce visualizations that incorporate intuitive interactions, overcoming the limitations of current LLMs' static outputs. Users should be able to interact with charts in ways that help them express their analytical intents naturally.

    \item \label{req:interpretion}
    \textbf{Understand the Intents of Multimodal Interactions.}
    The system needs to automatically and seamlessly integrate interactions from different modalities to precisely infer users' interaction and \mbox{analytical intents.}

    \item \label{req:accuracy}
    \textbf{Ensure the Correctness of the Visualizations.}

    Users need reliable analysis results free from errors caused by LLMs (e.g., hallucinations \cite{zhang2023siren}). By ensuring correctness in each stage of analysis, the system builds trust, minimizing inaccuracies and supporting user confidence in the analytical insights provided.
\end{enumerate}

\vspace{-0.7em}
\section{InterChat}

\begin{figure}[!t]
    \sidecomment{R1.3\\ R2.2 \ R3.8}
    \centering
    \revisionbox{
      \includegraphics[width=\linewidth]{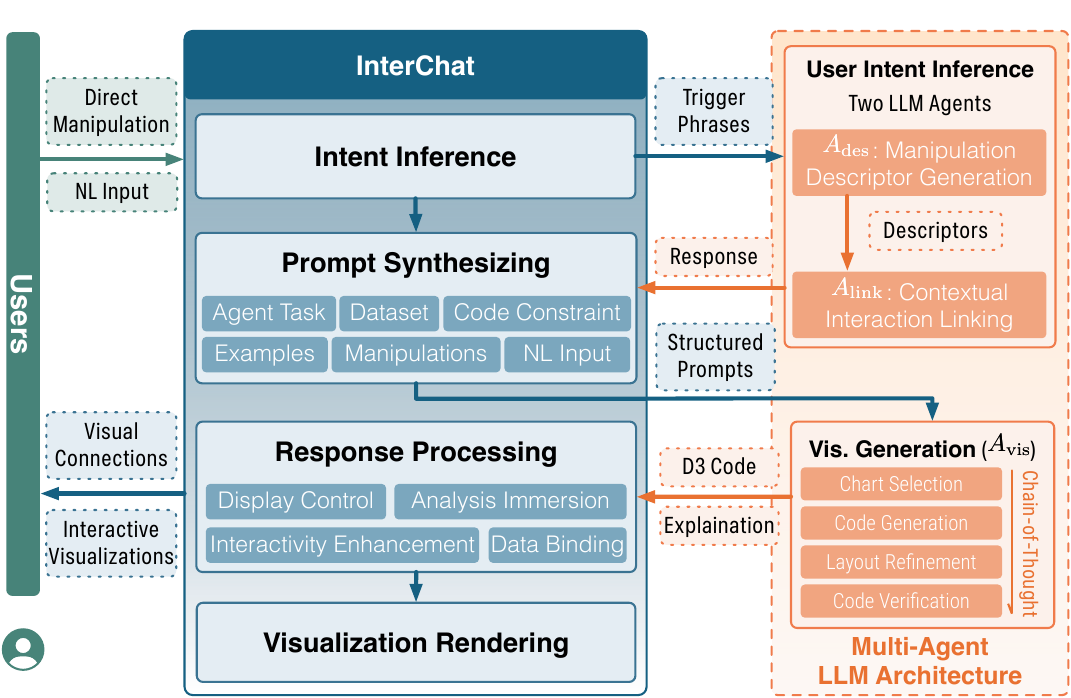}
    }
      \vspace{-0.5cm}
    \caption{
      The system workflow.
      \revisiontext{
        We use a multi-agent LLM architecture with two agents ($A_\mathrm{des}$, $A_\mathrm{link}$) for intent inference and one ($A_\mathrm{vis}$) for visualization generation. With our response processing module, users interact with visualizations to express analytical intents and inspect inferred intents through visual connections.
    }}
     
      \vspace{-1em}
    \label{fig:workflow}
\end{figure}

{
\renewcommand\UrlFont{\rmfamily}
\noindent
Following the design space and the design requirements identified, we develop InterChat to bridge users and LLMs in generative visual analytics.n{\sidecomment{R1.3 \ R2.2 \ R3.8}%
    The system employs a multi-agent LLM architecture, where two agents ($A_\mathrm{desc}$, $A_\mathrm{link}$) are used for intent inference and one agent ($A_\mathrm{vis}$) is used for visualization generation.

As shown in Figure \ref{fig:workflow},
\textbf{from users (left) to LLMs (right)}, InterChat enables users to upload and analyze their own datasets through multimodal interactions (\ref{req:generalizability})\sidecomment{R3.3}. This requires appropriate dataset content description and type definition for each attribute provided. Examples of these descriptions are available in Appendix I.
With $A_\mathrm{desc}$ and $A_\mathrm{link}$ (Section \ref{sec:intent_guessing}), InterChat builds connections among different modalities of user input and helps users visually confirm that their interactions have been correctly interpreted (\ref{req:interpretion}).
Then, InterChat synthesizes the user intents into structured prompts, which are sent into a Chain-of-Thought agent $A_\textrm{vis}$ to generate appropriate visualizations to meet the users' analytical needs with (\ref{req:accuracy}, Section \ref{sec:prompt_engineering}).
}

\textbf{From LLMs (right) to users (left)}, InterChat processes and renders D3.js visualization from LLMs' responses, performing display control, interactivity enhancement, and data injection for interactive visualization rendering with data context(\ref{req:interactivity}, Section \ref{sec:response_process}). With the generated results, users can refine analytical needs, correct misinterpretations, and further explore the data interactively.

}
\subsection{User Interface}
\label{sec:ui}

\begin{figure*}[t]
    \centering
    \includegraphics[width=0.92\linewidth]{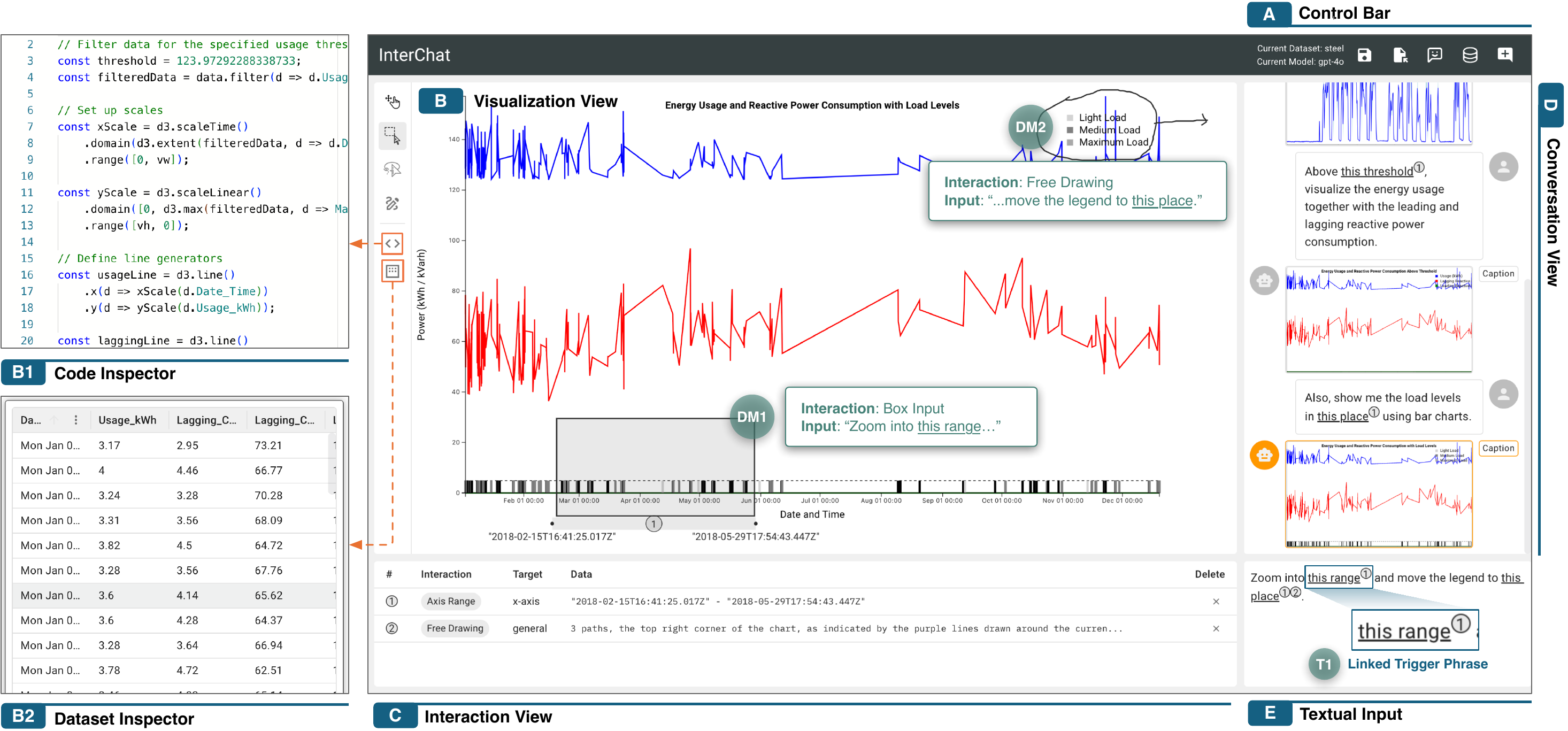}
    \caption{The user interface of InterChat system, showing energy consumption data from a steel manufacturing company. DM1 and DM2 refer to two Direct Manipulation operations user performed on the visualization, where DM1 involves selecting a temporal range using the Box Selection tool and DM2 involves conveying a chart reconfiguration intent of moving the legend to the top-right corner.}
    \label{fig:system}
    \vspace{-0.5cm}
  \end{figure*}

InterChat has five views, as shown in Figure \ref{fig:system}. On the right side are the Conversation View (D) and the Text Input View (E).
Similar to existing conversational UIs (e.g., ChatGPT \cite{openai2023chatgpt}), users can input text via both keyboard or microphone, and navigate through the conversation history.
\revision{\sidecomment{SR1}%
Each history entry includes the generated visualizations and the corresponding interactions.
Users can click on the history entry to review the corresponding results and interactions, edit the instructions or interactions to see the updated results.
The thumbnails for generated visualizations are displayed alongside the text responses.
}%
\revision{\sidecomment{R3.7}%
To initiate a session, users can explicitly provide instructions in the Input View on the analysis task or simply use open-ended queries like ``\textit{Show me the overview of the dataset}''.
}%
The top Control Bar (Figure \ref{fig:system}A) enables users to upload datasets, save or load chat histories, and switch between LLM models.


The Visualization View (Figure \ref{fig:system}B) renders the generated D3 code.
Users can perform interactions including \textit{click selection}, \textit{box selection}, \textit{lasso selection}, and \textit{free drawing}, with tools listed on the left.
These interactions cover all direct manipulations in our interaction space and are recorded in the Interaction View (Figure \ref{fig:system}C) in a tabular format.
Each row logs the direct manipulation type, manipulated visual elements, and related data items. For free drawing, inferred user intents from LLMs are displayed.

Additionally, InterChat includes a Code Inspector (Figure \ref{fig:system}B1),%
\revision{\sidecomment{R3.3}
allowing expert users to inspect or modify the generated code and developers to debug the system. The Code Inspector is an \textbf{optional }feature for advanced users knowing D3 and is hidden by default.
\sidecomment{4.10}
When edited, we apply a throttle of two seconds and re-render the canvas using the modified code, where all the previous interactions are preserved as long as the corresponding visual elements are kept.
}%
The Dataset Inspector (Figure \ref{fig:system}B2) displays the original dataset attributes and values in tabular form.

\subsection{Intent Inference for Multimodal Interactions}
\label{sec:intent_guessing}

InterChat supports four types of direct manipulations (\textit{click selection}, \textit{box selection}, \textit{lasso selection}, and \textit{free drawing}).

\subsubsection{Descriptor Generation}
\label{sec:descriptor_gen}

In this initial step, InterChat generates a concise \textit{manipulation descriptor} for each direct manipulation. This descriptor captures the key information of the manipulation, categorized into three cases:

\begin{enumerate}
    \item \textit{Interactions with Data-Related Visual Elements}. For interactions including click selection and lasso selection, the descriptor records the involved visual elements and associated data items. For example, a click on a bar in a bar chart is described as \textit{``user selected a bar elements, with data item: \{$\cdots$\}''}.
    \item \textit{Interactions with Axis Ranges}. With global X and Y scales, InterChat records the selected data range when used. For example, dragging on the canvas with the Box Selection tool is recorded as: ``\textit{selected data range on the x-axis: $[x_1, x_2]$ and y-axis: $[y_1, y_2]$}''. Here, we apply a $5\%$ threshold where selections smaller than this distance on the axis will not be recorded.
    \revision{
        \item \sidecomment{R3.5}\textit{Free Drawing}.
        Users can sketch on the visualization to directly express flexible intents.
        We first capture a screenshot of the canvas; this screenshot, along with NL inputs, are processed by agent $A_\mathrm{des}$, which employs a Vision LLM (GPT-4o by default) to interpret intent and generate a textual manipulation descriptor. For example, drawing an upward arrow on a line chart and stating, ``\textit{Find all segments with this trend},'' produces the descriptor: ``\textit{an arrow indicating a steady upward trend}.''
    }
\end{enumerate}
\revision{
The free drawing tool's functionality is largely dependent on the Vision LLM's interpretative capabilities. Our experiments show that it supports a variety of interactions, including but not limited to: describing trends with arrows, specifying regions for data selection, reconfiguring visualization layouts such as legend and axis location, and marking textual annotations for reference in natural language input. Example use cases can be found in Appendix II.
}
\subsubsection{Contextual Interaction Linking}
\label{sec:linking}

In the second step, InterChat connects the manipulation descriptors with the user's language input to refine intent inference. This is particularly important for multi-interaction queries.
\revision{\sidecomment{R1.4 \ R2.2 \ R3.8g}

It extracts and maps trigger phrases---continuous word sequences referencing direct manipulation, like ``this trend'' or ``the selected data''---to corresponding descriptors.
We use agent $A_\mathrm{link}$ to identify trigger phrases and the corresponding number of manipulations they reference. Our experiments show that LLMs can detect diverse trigger phrases, including pronoun-noun combinations (e.g., ``this trend'', ``those bars''), noun phrases (e.g., ``the area''), and adjective-noun phrases (e.g., ``the selected data'', ``the zoomed-in area'', ``the first selection''), etc.
}
The mapping process is based on the following rules, applied in descending order of priority:

\begin{enumerate}
    \item \textit{Order Matching}.
    Trigger phrases should match the descriptor order, assuming users describe actions sequentially. The number of trigger phrases must equal descriptors.
    \item \textit{Content Matching}. When a DM is mentioned multiple times, the number of trigger phrases and descriptors may differ. We instruct LLMs to link trigger phrases based on descriptor content. For example, ``these two time ranges’’ is linked to two Box Selections on the x-axis rather than a click on a circle in a scatter plot.
    \item \textit{Flexible Matching}:
    If the above strategies fail, we attempt to match the content of the trigger phrases with the descriptors. Unmatched descriptors are appended to the structured prompts and are handled by visualization generation agent $A_\textrm{vis}$.
\end{enumerate}

\noindent Successfully inferred intents are shown in the Interaction View (Figure \ref{fig:system}C), with trigger phrases in the text input box highlighted when users hover over the descriptors (Figure \ref{fig:system}T1).

\vspace{-1.5em}
\revisiontext{\subsection{Visualization Generation}}
\label{sec:prompt_engineering}

\noindent

\begin{figure}[t]
    \centering
    \revisionbox{
      \sidecomment{SR1\ R2.2}
    \includegraphics[width=\linewidth]{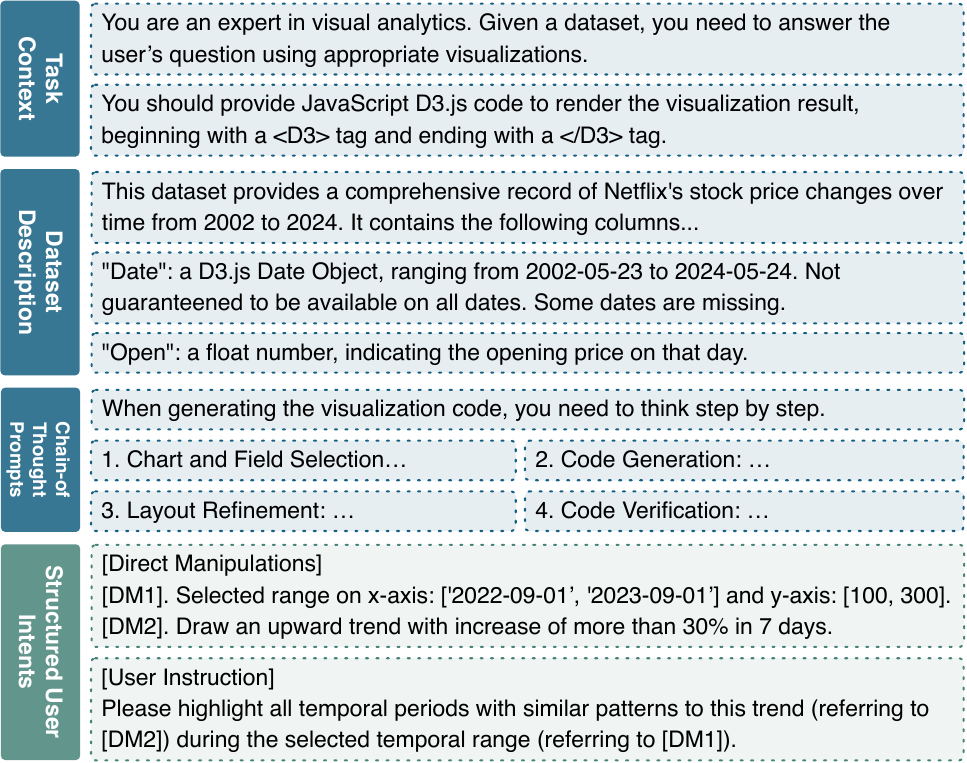}
    }
    \vspace{-1em}
    \caption{
      \revisiontext{
        Prompt structure for visualization generation. Examples shown here uses Netflix stock price data. We only display the first few lines of the prompts. Full content is available in Appendix I.
      }
    }
    \label{fig:prompt}
    \vspace{-3em}
\end{figure}

\revision{\sidecomment{SR1 \ R1.3 \ R2.2\ R3.8a}%
To generate accurate visualizations, InterChat uses the $A_\mathrm{vis}$ agent with structured prompts and Chain-of-Thought \cite{chainofthought} prompting. It organizes user intent into structured prompts and guides generation step by step. The prompt includes (Figure \ref{fig:prompt}):

\begin{enumerate}
    \item \textit{Task Context}:
    We instruct the agent to generate D3.js \cite{bostockD3} visualization code in an SVG context, specifying constraints on global variables, libraries, functions, data access patterns, forbidden APIs, and the expected response structure.
    \item \textit{Dataset Description}: We provide detailed dataset information, including source, meaning, attribute types, value ranges, and nullability. For datasets with many similar columns, we abbreviate attribute definitions, e.g., {\small\texttt{sensor{1-200}: float}}.
    \item \textit{Chain-of-Thought Prompts}: We instruct the generation process step by step, including: a) Chart and Field Selection, where the agent determines the appropriate visualization type and selects relevant data fields based on the user's analytical intent and the dataset; b) Code Generation, where the agent generates D3.js code under previously specified constraints; c) Layout Refinement, where the agent optimizes the layout, axis range, and color encodings for better visuals; d) Code Verification, where the agent verifies the correctness and fixes potential code bugs.
    \item \textit{Structured User Intents}: We organize the inferred intents into structured prompts, including direct manipulation descriptors and natural language inputs with their trigger phrases connected.
\end{enumerate}

\noindent Full prompts are in Appendix I.
\sidecomment{R3.5}%
Structured prompts were previously applied in DirectGPT \cite{directGPT} to enhance user interaction with LLMs by integrating direct manipulation design principles. We choose a similar approach but further incorporate automated interaction linking and vision LLM-based intent interpretation.
}

We chose D3.js for two reasons. First, its widespread use in large language models' training data improves their familiarity with D3, leading to better generated code. Second, D3 offers a balance between simplicity and flexibility, enabling the creation of high-quality visualizations.
\revision{%
\sidecomment{R3.8b}
Compared with higher-level tools like Vega-lite \cite{arvind2017vega}, D3 offers finer control over elements and enables direct manipulation by injecting JavaScript into the generated visualization. The data-binding mechanism significantly helps extract selection targets for direct manipulation.
In our implementation of using GPT-4o, reliably supported visualization types include bar charts, line charts, scatter plots, histograms, pie charts, area charts, and heatmaps.
}
While using lower-level specifications such as directly generating SVG elements, the significantly increased complexity can easily lead to more error-prone code generation.

\vspace{-1em}
\subsection{Response Processing and Visualization Rendering}
\label{sec:response_process}

\noindent

\noindent
The output of $A_{\textrm{vis}}$ contains a D3.js code snippet and some explanation of the user's request.
To facilitate interactive analysis, InterChat then performs the following steps to process the responses:

\textbf{Analysis Immersion}.
Considering users' preference of immersing themselves in the analysis process without being distracted by code, we extract code from our specified {\small\texttt{<D3></D3>}} tag and hide it by default. For the textual response, we require agents to only explain visual encodings and the rationale of task execution. The logic of the code is only shown in the code comments. When code modification is needed, users can click the Code Inspector (Figure \ref{fig:system}B1) to view and edit the code, where the code is displayed in a Monaco Editor \cite{microsoft2023monaco} and updated in real-time.

\textbf{Display Control}.
Since we use multiple containers with different sizes to display visualization, we enforce LLMs to explicitly use three global variables: 1) the root container {\small\texttt{svg}}, 2) specified width {\small\texttt{vw}}, and 3) height {\small\texttt{vh}}. By injecting different root containers and viewport sizes, InterChat can dynamically adjust the position and size of visualizations to fit the desired context.

\textbf{Data Binding}.
\revision{\sidecomment{R3.6}%
InterChat avoids providing detailed dataset values to LLMs during the generation process for two reasons:
(1) including full data values exceeds the practical input length supported by LLMs; and (2) LLMs are not well-suited for computations or reasoning over large numerical sequences.
}%
Therefore, only a concise description of the data schema is provided to the LLMs, including the name, data type, and attributes of each field.
All computations and data handling are executed locally. Upon receiving the d3.js code generated by the model, we inject the dataset as a global variable {\small\texttt{data}}, represented as a JavaScript object array containing full data values.
\revision{\sidecomment{R4.6}%
Necessary preprocessing tasks, such as handling missing data or correcting malformed date strings, can be performed in two ways: users can optionally specify mapping functions in JavaScript within the system, while the generated code also includes essential preprocessing steps required for visualization.
}%

\textbf{Interactivity Enhancement}.
\label{sec:interactivity_enhancement}
InterChat incorporates two key improvements to the generated code to enhance the interactivity of visualizations.
First, to track user activity, all visualizations are required to use a single pair of global X and Y linear scales. If multiple subplots are included, they must share unified scales. These global X and Y scales are always returned by the generated code, enabling InterChat to track x-axis and y-axis values when users interact with the visualization (e.g., hovering, clicking, or drawing).
Second, to support contextual interaction linking (Section \ref{sec:intent_guessing}), InterChat binds the full data item to its corresponding visual element created by D3.js. Technically, this is implemented by modifying the D3.js code to replace all chains of {\small\texttt{.data().enter().append()}} calls to {\small\texttt{.attr('data', d =$>$ d.toString())}}. This ensures that each visual element retains a reference to its associated data.

\section{Evaluation}

To \revision{\sidecomment{R1.5}%
demonstrate the effectiveness and usefulness of leveraging multimodal interactions for visual analytics tasks, we present two usage scenarios utilizing datasets from Kaggle, conduct a user study with ten participants, and collaborate with industry experts to assess the system's usability in industrial settings.
}%

\subsection{Usage Scenario I: Netflix Stock Price Analysis}

\begin{figure*}[t]
  \centering
  \includegraphics[width=1\linewidth]{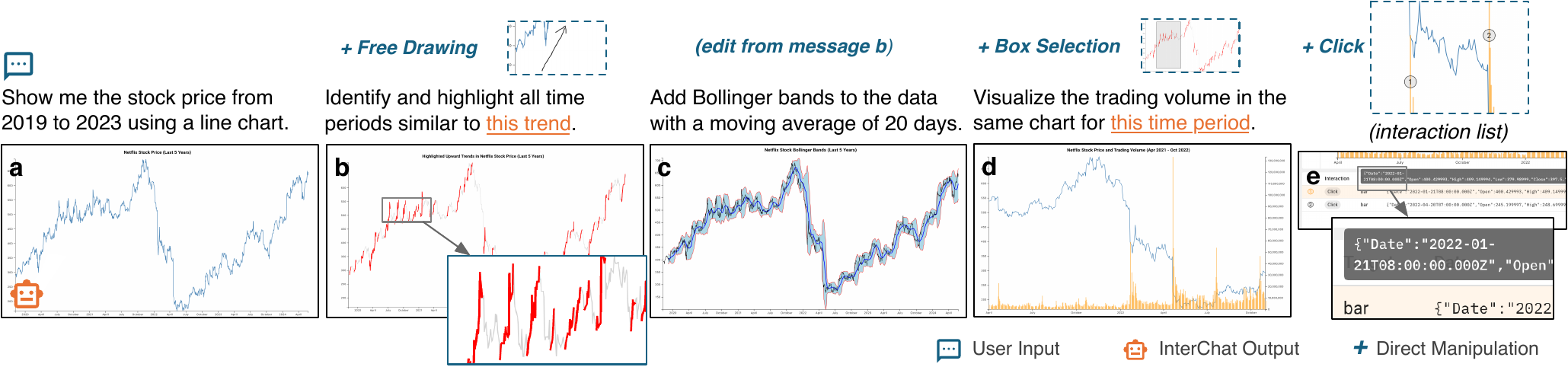}
  \caption{Usage scenario I: Netflix stock price data exploration and trend analysis. The user starts with a line chart (\bf{a}), instruct the system to create Bollinger bands (\bf{c}), uses free drawing tool to search for a specific trend (\bf{b}), and employ box selection to select a specific time period (\bf{d}). Our data binding mechanism allows users to inspect corresponding data items by clicking the chart elements (\bf{e}).
  }
  \label{fig:case-stock}
  \vspace{-0.5cm}
\end{figure*}

The \textit{Netflix Stock} dataset\revision{\sidecomment{R3.8c}\cite{netflixStock}} contains stock prices and trading volumes for Netflix from 2002 to 2024, with $5,540$ time steps with seven columns: {\small\texttt{Date}}, {\small\texttt{Open}}, {\small\texttt{High}}, {\small\texttt{Low}}, {\small\texttt{Close}}, {\small\texttt{AdjClose}}, and {\small\texttt{Volume}}, accessible via Dataset Inspector (Figure \ref{fig:system}B2). In this scenario, we aim to perform exploratory data analysis to reveal historical stock price trends and understand trading behaviors.

\bf{Identifying Historical Trends}:
We began by instructing the system, via natural language input, to generate a line chart displaying the stock price over the last five years. The system accurately interprets our intent and renders the visualization in the Visualization View promptly (Figure \ref{fig:case-stock}a). We then used the Free Drawing tool to sketch an upward trend on the chart and instruct the system to highlight all the upward trends similar to the one we sketched (Figure \ref{fig:case-stock}b). The Intent Inference module interprets the user-defined trend as \textit{"a steep and consistent upward trend of more than 100 USD in a short period of no more than 7 days"}, and updates the visualization accordingly with matching areas highlighted.
To further examine price volatility, we added Bollinger Bands with a 20-day moving average to the chart, enabling us to observe price movements within the bands in greater detail (Figure \ref{fig:case-stock}c).

\textbf{Understanding Trading Behaviors}:
The historical stock price visualization reveals a significant decrease starting in late 2021, followed by a partial recovery in mid-2022---a trend likely reflecting intensified competition in streaming media affecting Netflix's market share and stock value.
To further investigate trading behaviors during this period, we used the Box Selection tool to focus on the period, instructed the system to zoom into this range, and also visualized the trading volume in the same chart using different axes (Figure \ref{fig:case-stock}d). We could easily observe two peaks on January 21, 2022, and April 20, 2022, which are also the days marked by the most significant stock price drops. By clicking these volume peaks, we could examine the open and close prices for these days in the Interaction List, enabled through the interactive data binding mechanism (Section \ref{sec:interactivity_enhancement}). A quick review of the news on these two days confirms that these peaks followed Netflix's release of its quarter earnings, which highlighted a notable decrease in subscriber growth and revenue \cite{NetflixSharesSlide2022}.

\subsection{Usage Scenario II: Steel Manufacturing}
\label{sec:case-steel}

The \textit{Steel Manufacturing} dataset\revision{\sidecomment{R3.8c}\cite{steelManufacturing}} records energy consumption from a steel company manufacturing steel products. It has $35,041$ time steps at 15-minute intervals, with attributes such as energy usage, reactive power (leading/lagging), $\textrm{CO}_2$ emissions, and load status. Our objective is to explore energy consumption patterns and correlations between reactive power and $\textrm{CO}_2$ emissions.

\begin{figure}[thbp]
    \centering
    \vspace{1em}
    \includegraphics[width=\linewidth]{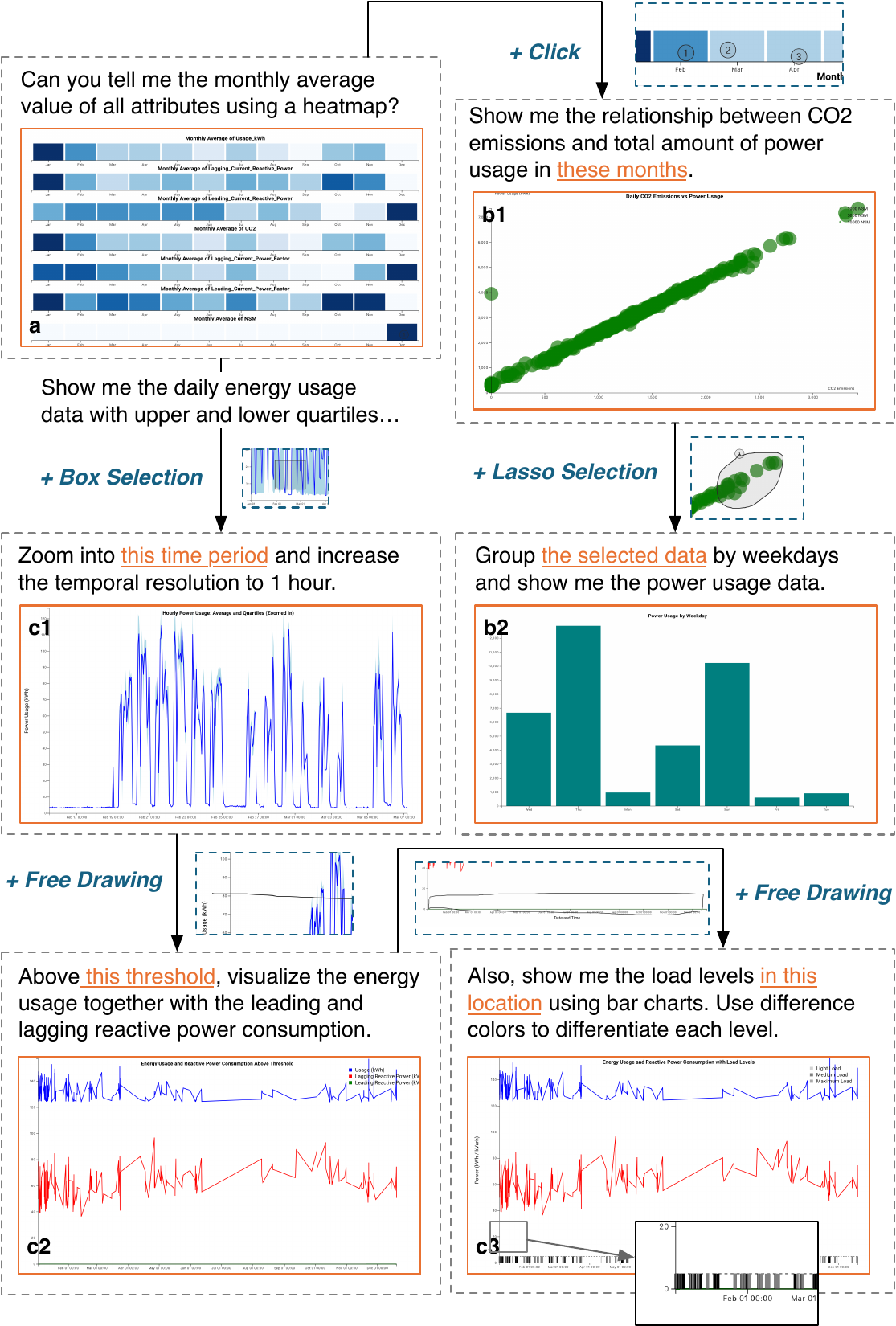}
    \caption{Usage Scenario II: visual analytics of energy consumption data for a steel manufacturing company. We start with a heatmap displaying the overview for all attributes in the dataset (\bf{a}) and start two conversations (\bf{b1} - \bf{b2}, \bf{c1} - \bf{c3}). The lasso selection tool is particularly helpful for isolating data points for scatter plots (\bf{b1}). The free drawing tool is used for multiple purposes, including specifying a threshold (\bf{c1}) and reconfiguring the visualization (\bf{c3}).}
    \label{fig:case-steel}
\end{figure}

\textbf{Exploring Energy Consumption Patterns}:
To start, we used a heatmap to visualize the value for all attributes in the dataset aggregated by month, providing an overview of the energy consumption information (Fig \ref{fig:case-steel}a). In this visualization, darker colors indicate higher values. The heatmap reveals that energy consumption peaks in January and February before steadily declining through the spring months. Next, we clicked on the rectangles to select three months to examine the correlation between $\textrm{CO}_2$ emissions and energy usage during this period. The intent inference module automatically matches \textit{these months} to the user selection, generating a scatter plot with energy usage on the x-axis and $\textrm{CO}_2$ emissions on the y-axis (Figure \ref{fig:case-steel}b1). The scatter plot reveals a linear relationship between these two attributes, except for a single day in February with zero $\textrm{CO}_2$ emissions, possibly due to sensor malfunction. We further employed the lasso selection tool to isolate specific data points and instructed the system to visualize energy usage per weekday for the selected period (Figure \ref{fig:case-steel}b2). This chart highlights peak energy consumption on Tuesdays, followed by Thursdays.

\textbf{Reconfiguring Visualizations}:
Next, we aimed to create a compiled visualization that examines leading and lagging reactive power over time, incorporating load status levels as a reference. We generated a line chart displaying daily power usage along with the upper and lower quartiles. Using the box selection tool, we zoomed into a specific time range and increased the temporal resolution to 1 hour (Figure \ref{fig:case-steel}c1). We then applied a threshold by drawing a line directly on the chart, filtering out hours with low energy consumption, and focusing on periods with high energy usage(Figure \ref{fig:case-steel}c2). Finally, we sketched a rectangle on the bottom of the chart instructing the system to add a subplot displaying the load status levels for the selected period (Figure \ref{fig:case-steel}c3). The system interprets the drawing as \textit{"a rectangle at the bottom of the chart, right above the x-axis with a height of 10 kWh"}, and generates the desired visualization. The resulting chart presents reactive power and load status in two subplots, enabling the observation of correlations between the attributes and the impact of reactive power on the load status.

\vspace{-1em}
\subsection{User Study}

\subsubsection{Methodology}

We conducted a user study with ten participants (P1 - P10) from the visualization community, averaging 4.2 years of experience in visual analytics. All were familiar with conversational language models.
\revision{\sidecomment{R1.6}%
The study aims at determining whether multimodal interactions improved the efficiency and accuracy of visual analytics tasks compared to language-only input, and collecting qualitative feedback on usability and overall user experience.
\revision{\sidecomment{R3.8e}%
Every participant used InterChat for ten analytical tasks involving the \textit{Netflix Stock} and \textit{Number of Deaths} datasets,
}%
which were general-audience-friendly and less domain-specific.
}%
Here, \textit{Number of Death} data contains mortality data across different countries, age groups, and time.

Half the tasks addressed simple intents (e.g., ``\textit{Find the monthly average trading volume over three years}''), while the other half involved complex, multi-step intents (e.g., ``\textit{Identify countries with the largest gender disparity in death rates}'').
The study was conducted online with screen recording, starting with a brief system introduction and five minutes of free interface exploration.
To minimize bias, we applied counterbalanced task orders and randomized task conditions, ensuring equal testing of each task with multimodal and language-only settings (Detailed in Appendix III). Each task had a time limit of five minutes.
The metrics collected included task completion time, retries for language input, visualization accuracy, and the correctness of interpreted intents (classified as correct, partially correct, or incorrect).
After the study, we further conducted 15-minute semi-structured interviews to collect qualitative feedback on usability and overall user experience.

\subsubsection{Results}

\begin{figure}[htbp]
    \centering
    \includegraphics[width=\linewidth]{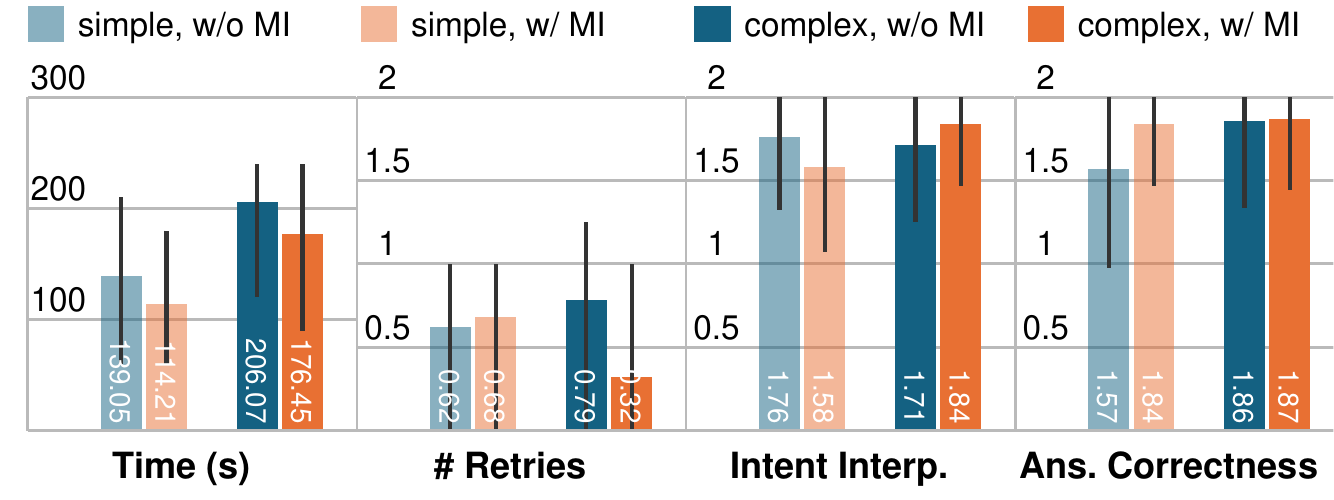}
    \caption{The compiled results for the user study. Here, "simple" indicates simple tasks with only analytical intents, and "complex" indicates complex tasks with multiple interactions and analytical intents. MI indicates Multimodal Interaction. \# Retries indicates the number of language prompts made by the participants. For intent interpretation and answer correctness, each task is graded as 0 (incorrect), 1 (partially correct), or 2 (correct).}
    \label{fig:user-study}
    \vspace{-1em}
\end{figure}

Figure \ref{fig:user-study} shows the compiled results.
Multimodal interactions reduced task completion times by $18\%$ for simple tasks and $15\%$ for complex tasks, with the average number of retries dropping significantly from $0.79$ to $0.32$.
This improvement stems from allowing direct interactions to reduce repetitive NL inputs for describing intents or selection criteria.
Efficiency gains were further reflected in improved visualization correctness and intent accuracy for complex tasks.
However, for simple tasks, intent accuracy was slightly lower with multimodal interactions, likely due to occasional misinterpretations introduced by the added complexity.

\textbf{Usable and Flexible System Workflow}:
Participants responded positively to the system's multimodal interaction, with 7 out of 10 finding \textit{selection} and \textit{filtering} tools most helpful. These tools support exploratory data analysis by enabling focus on specific components or ranges. Box and Lasso selection provide intuitive filtering without ambiguous language input, especially for scatter plots. Participants also valued the flexibility of the Free Drawing tool: P3 sketched subplot layouts, P4 annotated reference line locations, P5 set axis ranges, and P10 marked a ``target'' for the legend. These examples highlight its expressiveness in capturing user intent.

\textbf{Simplifying Complex VA Tasks}:
When users are uncertain of exact requirements and need progressive exploration, multimodal interactions offer flexible and intuitive means to delve into the dataset. For example, to compare open and close prices on days with high trading volume, P5 generated a bar chart for trading volume, selected the bars with high volume, and instructed the system to generate a line chart. Conversely, for well-defined tasks like single-intent queries, participants often preferred direct language input, perceiving multimodal interactions as unnecessary.

\textbf{Suggestions}: Participants suggested several improvements. P4 recommended expanding interactions to include drag-and-drop functionalities for moving data fields from the Data Inspector to the Visualization View. P9 proposed incorporating predictive capabilities based on conversation context and interaction history to expedite requirement specification. P6 expressed interest in enhancing the interactivity of visualizations with features like interactive tooltips on hover or direct color configuration for visual elements.

\subsubsection{Expert Feedback}

We also collaborated with two industry experts from the manufacturing sector to evaluate the system's functionality in industrial scenarios. The experts conducted exploratory analyses using the \textit{Steel Manufacturing} dataset and assessed InterChat's capabilities within their internal workflows.
\revision{\sidecomment{R3.8d}%
They highlighted the system's flexibility in generating a wide range of fundamental charts, potentially replacing the functionality of manually dragging and dropping data fields and setting filter conditions or chart parameters in Tableau.
}%
The experts appreciated the multimodal interaction features for reconfiguring visualizations, particularly for selection purposes.
However, they also expressed the need for computational capabilities to support functions such as linear regression or correlation analysis. Due to data privacy concerns, we provided the model with only the data structure rather than raw values, limiting the ability to perform such calculations directly.
They suggested that the workflow can be further enhanced by displaying multiple subplots simultaneously and incorporating interactive coordination between different views.
This feedback points to a promising direction for future work, such as enabling the system to generate Python-based computational scripts to address these requirements.

\section{Discussion}

\subsection{Implications of Multimodal Interactions in Generative Visual Analytics}

\bf{Reducing Cognitive Load}:
Multimodal interactions significantly alleviate the cognitive load required to express analytical intents, particularly for complex tasks involving multiple interactions. By enabling users to directly manipulate visualizations---specifying selection criteria or intuitively sketching intents—they circumvent the need to compose detailed language inputs. This direct engagement allows users to focus on data exploration without formulating precise verbal prompts. Examples include selecting data points in a scatter plot, delineating date ranges in a time series chart, or illustrating patterns in a line chart. Our user study reflects this benefit, showing a significant reduction in retries for complex tasks when utilizing multimodal interactions, indicating smoother analytical workflows and fewer interruptions.

\bf{Enhancing Credibility}:
Natural language-based visualization generation often operates as a black box, limiting users' insight into how the system interprets their intents and potentially leading to inaccurate or misleading visualizations due to LLM hallucinations.
In our user study, participants frequently expressed doubts about the correctness of visualizations generated through NL-only input,
\revision{\sidecomment{R4.2}%
primarily due to the end-to-end nature of the process, which lacks intermediate visual outputs for verifying the system's intent interpretation.
Conversely, multimodal interactions provide transparency by allowing users to interactively filter and reconfigure charts and provide explicit visual feedback. This visibility enables the timely correction of misinterpretations through prompt or interaction edits, thereby enhancing trust and credibility for the results.
}%
Future iterations can explore integrating semi- or fully-automated verification mechanisms with MIs to further enhance confidence.

\bf{Facilitating Collaborative Data Analytics}:
The expressiveness of multimodal interactions is particularly advantageous in collaborative environments such as online meetings or shared workspaces. This potential is evidenced by positive feedback on the Free Drawing tool, which, unlike language-only inputs, enables users to sketch intents directly on the canvas. In collaborative settings, multiple users can annotate, highlight areas of interest, or define selection criteria on visualizations, with changes instantly visible to all participants. This capability transforms the system into a collaborative platform, enabling synchronized discussions and decision-making, thereby establishing generative visual analytics systems as essential tools for distributed teamwork scenarios.

\vspace{-0.5em}
\revision{
\subsection{Failure Analysis}
\sidecomment{SR3 \ R3.5 \ R3.8a \ R3.8e \ R4.1}%
We analyzed 100 task executions collected during the user study and identified two main types of failures:

\textbf{Failed intent interpretation:} In 26 cases, the system misinterpreted user intent, leading to incorrect outcomes. Common issues included layout misinterpretation (9 cases), such as failing to recognize the user's preference for placing multiple charts side by side; ambiguity in trend recognition when using the free drawing tool (4 cases), where the system overgeneralized upward or downward trends and could not properly highlight the requested segment; and inaccuracies in detecting selected objects when using the drawing tool (4 cases), failing to recognize the selected objects accurately.

\textbf{Failed visualization generation:} In 4 cases, the generated D3.js code failed to execute due to syntax or runtime errors, including two cases where the system attempted to access undefined variables and two cases where non-existent D3 functions required third-party libraries. Additionally, in 12 cases, although the generated visualizations were functional, they contained issues, including incorrect axis placement (6 cases), overlapping legends and chart elements (3 cases), and inappropriate color encodings that did not align with the semantic meaning of the data (3 cases).

To recover from these failures, users can utilize the edit feature to revert to a previous conversation state while retaining the interaction context. From our observation, after at most three retries, users could typically obtain a satisfactory result. Many of these failures could be mitigated by providing additional constraint prompts to guide the model's interpretation. A simple retry could also be effective in many cases. As LLMs continue to advance, we expect improvements in intent inference and visualization accuracy.
}

\subsection{Limitation and Future Work}

\revision{\sidecomment{R1.2 \ R4.2}%
\bf{Latency and Accuracy:} While our approach demonstrates flexibility across various tasks, its latency and accuracy remain constrained by LLMs. Users experience longer wait times for visualization outputs, which can disrupt workflow and lead to loss of context. Additionally, LLM hallucinations may result in incorrect transformations or visualizations, making selections less precise compared to traditional menu-based interfaces. Although multimodal interactions enhance user trust, the system currently lacks mechanisms to verify output correctness. Future work could focus on developing methods to detect and mitigate LLM hallucinations or training a more efficient Teacher-Student model specifically for visual analytics tasks to improve accuracy and efficiency.
}%

\bf{Data Limitations and Computational Capabilities:}
\revision{%
Our current pipeline provides only the metadata of the data attributes into the language model, limiting its computational capabilities to simple data transformations and aggregation such as filtering, grouping, and sorting. Complex transformations and computation that require direct access to raw data are not supported.
\sidecomment{R3.5}}%
Given our extensible multi-agent LLM architecture, future work could explore incorporating some data values into the conversation context, enabling more sophisticated data transformations and computational tasks, such as correlation analysis \mbox{or outlier detection.}

\bf{Scope of Interaction Modalities:}
Although our research focuses on interactions on traditional 2D screens, more complex modalities---such as advanced touch gestures, voice commands, and movement-based interactions---remain underexplored.
Emerging technologies like Augmented Reality (AR), Virtual Reality (VR), wearable devices, and mobile platforms introduce new opportunities for enriching multimodal interactions in VA.
\revision{\sidecomment{R2.3 \ R4.2}%
Meanwhile, the integration of multimodal interaction with conventional actions (e.g., brushing, linking) could be further explored to enhance the interactions' expressiveness and flexibility.
Future work could investigate how these modalities can be integrated with existing approaches, potentially enabling more efficient, interactive, and intuitive analytical experiences across various environments.
}%

\vspace{-0.3em}
\section{Conclusion}

This work introduced InterChat, a multimodal generative visual analytics system that integrates natural language and direct manipulations to enable precise expression of analytical and interaction intents. Guided by insights from pilot brainstorming sessions, we identified key design spaces and requirements that shaped the system’s development. InterChat advances generative visual analytics through effective prompt design, intent inference, and response processing.
Demonstrated through diverse usage scenarios, a user study, and expert feedback, the system showcases efficiency and effectiveness. InterChat establishes a solid foundation for future advancements in multimodal analytics systems.
\vspace{-0.3em}

\section*{Acknowledgments}
We thank the participants of both user studies for their valuable feedback. This work was supported in part by the U.S. National Science Foundation under Grant III-2427770 and by a Bosch Research Award.

\bibliographystyle{eg-alpha-doi}
\bibliography{main}



\section*{Supplementary material}

# InterChat: Enhancing Generative Visual Analytics using Multimodal Interactions

## Appendix

### I. Prompt Design

*A. Prompts for Visualization Generation Agent $A_{\text{vis}}$*

These prompts define the system behavior and will be provided at the beginning of each conversation. The system prompts are as follows:

```
 1  [Task Context]
 2  You are an expert in visual analytics. Given a dataset, you need to answer the user's questions using appropriate visualization and text annotations.
 3  You should provide a visualization that best represents the data and answers the user's questions.
 4  You should provide JavaScript D3.js code to render the visualization result, beginning with <D3> and ending with </D3>.
 5  The chart will be rendered in an svg element with a view box from (0,0) to (vw,vh), the instance of which is stored as a variable named svg.
 6  Note that you need to specify the color of each element and add necessary elements like legend and axis.
 7  Note that I will not preserve any variables you defined in the previous dialogue. You need to define every necessary variable before you use it in each code block.
 8  You must explain the visualization without explaining the code since the users have no knowledge of D3.js.
 9  You must add a title on the top of the chart using bold text.
10  You must return [xScale, yScale] at the end of the code block, before the </D3> (for each, null represents N/A). Only one xScale and one yScale should be used. If multiple subplots are needed, you should use a parent container with a single x /y axis.
11  You must try your best to make the code correct and handle all undefined.
12  You should not redefine vw, wh, or the data variable.
13  You should not use d3.nest() or d3.rollup() or d3.group() functions.
14  You must not output '''javascript''', <html>, or <svg> tags in the results.
15  You must generate d3 visualization as a response to the user query, no matter if it is analytical, exploratory, or explanatory.
16
17  [Dataset Description]
18  (See Section I.C)
19
20  [Procedure]
21  When generating visualization code, you need to think step by step.
22  1. Chart and Field Selection: you need to analyze the user's analytical intent and the dataset characteristics to determine the most appropriate chart type and select relevant data fields.
23  2. Code Generation: write the complete D3.js code following the constraints mentioned above. Ensure the code is correct and readable.
24  3. Layout Refinement: Optimize the chart layout, including axis ranges, color encodings, and placement of legends and title for effective visual communication.
25  4. Code Verification: Validate the generated code for correctness and resolve any potential issues. Commonly encountered issues include access to undefined variables, incorrect axis ranges, unexpected data types, and inappropriate library usage.
26
27  For example, when users want to inspect the change in stock price, you need to generate a line chart with date field as the x-axis, and stock price as the y-axis with a chart title: "Netflix Stock Price Change".
28
29  [Direct Manipulations]
30  The user's direct manipulations are as follows:
31  [DM1]. {DM_DESCRIPTOR[0]}
32  [DM2]. {DM_DESCRIPTOR[1]}
```

```
33  ...
34
35  [User Instruction]
36  The user's language input is: {NL_INPUT}.
```

## B. Prompts for Intent Inference of Free Drawing $A_{\text{intent}}$

These are the system prompts used to generate manipulation descriptors from users' free drawing input. `NL_INPUT` and `TRIGGER_PHRASES` are placeholders for the user's natural language input and the trigger phrases extracted from the input, respectively.

```
1   [Task Context]
2   You are an expert in visualization analytics.
3   You will be provided with an image containing D3 visualization results and users' drawings.
4   The user drew some purple sketches on the visualization and said, "{text}". You should generate a descriptor for the users'
    direct manipulation in detail.
5   You should make the inference step by step and write your conclusion in a <Conclusion></Conclusion> block.
6   You should always put answers in the <Conclusion></Conclusion> block, starting with "{TRIGGER PHRASES} means".
7   If you are not sure about the interaction, response <Conclusion>UNKNOWN</Conclusion>.
8
9   [Examples]
10  For example, users may draw an arrow on a line chart to represent a trend. You need to specify whether is it an increasing
    trend, a decreasing trend, or an increasing-flat-decreasing trend and response:
11  <Conclusion>The user specified an upward trend of 30% in no more than 7 days.</Conclusion>
12
13  [User Instruction]
14  The user's input: {NL_INPUT}
15  Trigger phrases: {TRIGGER_PHRASES}
16  What does the user mean?
```

## C. Prompts for Contextual Interaction Linking Agent $A_{\text{link}}$

These prompts are used to extract trigger phrases from users' natural language input. A trigger phrase refers to a continuous sequence of words that reference direct manipulation, such as "this trend" or "the selected data". For multi-object references, the agent also extracts the quantity of manipulators being referred to.

```
1   [Task Context]
2   You are an expert in visualization analytics.
3   You are responsible for extracting trigger phrases in a multimodal visual analytics system where users interact with
    visualizations using natural language combined with direct manipulations (free drawing, box/lasso selection, and clicking).
4
5   Your task is to identify continuous words that refer to direct manipulation. Example trigger phrases include this trend, that
    circle, the selected data, this axis, those points, and these bars.
6
7   For each identified trigger phrase, extract:
8   - content: the exact phrase text
9   - quantity: number of objects being referred to
10
11  Note that trigger phrases must be continuous text sequences.
12  Only return results in this JSON format: { content: string; quantity: number;}[]
13  Return multiple trigger phrases in order of occurrence.
14  Return [] if no trigger phrases are found.
15
16  [Example]
17  Input: "What are the top 10 most expensive cars with these two colors? Please rank them using a horizontal bar chart."
18
```

```
19  Ouput:
20  [
21    {
22      "content": "these two colors",
23      "quantity": 2,
24    }
25  ]
26
27  [User Instruction]
28  The user's input: "{NL_INPUT}". What trigger phrases are included here?
```

## D. Dataset Description

These prompts define the structure and content for each column of the dataset, facilitating visualization code generation.

*a). Netflix Stock*

```
1   This dataset provides a comprehensive record of Netflix's stock price changes over time. It includes the following columns:
2
3   "Date": a D3.js Date Object, ranging from 2002-05-23 to 2024-05-24. The dataset is not always available on all dates.
4   "Open": a float number, indicating the opening price on that day.
5   "High": a float number, indicating the highest price on that day.
6   "Low": a float number, indicating the lowest price on that day.
7   "Close": a float number, closing price on that day.
8   "Volume": a float number, trading volume of the day.
9
10  You do not need to parse the date again as it is already a D3.js Date Object.
11  The range for stock price and trading volume are constantly changing. You need to recalculate the max and min values for
    different time periods.
12  You can directly use the variable named data to access the dataset, which is in the format {Date: string, Open: number, High:
    number, Low: number, Close: number, AdjClose: number, Volume: number}[]',
```

*b). Steel Manufacturing*

```
1   This dataset contains the steel manufacture data, including the following columns:
2
3   "Date_Time": a D3.js Date object, representing the date and time when the data was recorded.
4   "Usage_kWh": a float number, Energy usage in kilowatt-hours (kWh), representing the amount of electricity consumed during a
    specific period related to steel production processes or facility operations.
5   "Lagging_Current_Reactive_Power": a float number, reactive power consumption in kilovolt-amperes reactive hour (kVarh). Reactive
    power is essential in steel industry equipment for tasks such as magnetization and induction heating.
6   "Leading_Current_Reactive_Power": a float number, Similar to lagging current reactive power but likely refers to leading
    reactive power consumption, which can occur in capacitive loads.
7   "CO2": a float number, Carbon dioxide emissions in metric kgs of CO2. This could indicate the environmental impact of energy
    consumption in steel production.
8   "Lagging_Current_Power_Factor": a float number, measures the efficiency of electrical power usage. A lagging power factor
    indicates inefficient use of electricity, which could signify inefficiencies in certain steel production processes or equipment.
9   "Leading_Current_Power_Factor": a float number, Similar to lagging power factor but refers to leading power factor, which
    indicates more efficient use of electricity.
10  "NSM": an integer number, indicating the number of seconds from midnight.
11  "Weekstatus": a string, could be Monday, Tuesday, ... Sunday, etc, indicating day of week that the data was recorded.
12  "Load_Type": a string, one of {"Light_Load", "Medium_Load", "Maximum Load"}.
13
14  You can access the entire dataset via the "data" variable.
15  You need to sort the data by "Date_Time" before performing any operations. You must not parse the date again since it is already
    a D3.js Date object.
```

### c). Number of Death

```
1   This dataset records the number of deaths in different countries, different age groups, genders, in different decades. It
    contains the following columns:
2
3   "Year": a integer number, one of {1970, 1980, 1990, 2000, 2010}.
4   "Age": a string, one of 0-6 days, 7-27 days, 28-364 days, 1-4 years, 5-9 years, 10-14 years, 15-19 years, ..., 75-79 years, and
    80+ years. You do not have to include all age groups in the visualization unless explicitly required.
5   "Sex": a string, one of "Male", "Female".
6   "Deaths", a number, representing the number of deaths in that year, age group. The range constantly changes for different time
    periods and in different countries, so you need to recalculate the max and min values for different time periods.
7   "DeathRate", a number, representing number of death per 100,000 people.
8
9   You can access the dataset via a variable named data, which is in the format {Country: string, Year: number, Age: string, Sex:
    string, Deaths: number, DeathRate: number}[].
```

### d). Cars Manufacturing

```
1    This is a dataset that records car manufacturing status, including the following columns:
2
3    "Car ID": a number, sequential, representing the unique identifier of each car.
4    "Brand": a string, representing car brand, such as Toyota, Honda, Ford, Chevrolet, etc.
5    "Model": a string, representing car model, such as Camry, Civic, Cruze, etc.
6    "Year": a number, representing the manufacturing year of the car.
7    "Color": a string, representing the color of the car.
8    "Mileage": a number, representing the total mileage of the car.
9    "Price": a number, representing the price of the car in dollars.
10   "Location": a string, representing the location of the car, such as New York, Los Angeles, etc.
11
12   You can access the entire dataset via the "data" variable.
```

### e). UCI-SECOM

```
1    This is the UCI SECOM dataset from the semiconductor manufacturing process, containing the monitoring signals/variables
     collected from sensors or process measurement points.
2    Each row denotes a product instance, including the following columns:
3
4    "Time": a date string in the format of "%Y-%m-%d %H:%M:%S".
5    "Pass/Fail": boolean, indicates if this single product is passed or not. If a product is passed, its value is true;
6    589 sensor reading columns with column name: an integer number ranging from 1 to 589.
7
8    Some sensor readings are missing and will be stored as NaN. You should always replace them with 0.
9    The value range of different sensors is different.
10   The data is not ordered. So you will first ned to sort the data by the "Time" column before performing time-related operations.
11   You can directly use the variable named data to access the dataset, which is in the format {"Time": string, "Pass/Fail":
     boolean, "1": number, "2": number, ..., "589": number}[]',
```

### E. Full Example including User Prompts

The user prompts contain the user's natural language input, descriptors for direct manipulation, and the selected data (if applicable). We provide a chat record example, including the user's prompt, the system's response, and the generated D3 code for visualization below. For readability, we formatted the JSON data to allow multi-line strings and comments.

```JSON
1   [
2     {
3       "role": "system",
4       "content": "..." // See the system prompts in Appendix I.A
```

```
5      }
6      {
7        "role": "user",
8        "content": "Show me the stock price in the last five years.",
9        "actions": [],
10       "instruction": {
11           "text": "Show me the stock price in the last five years.",
12           "references": []
13       }
14     },
15     {
16       "role": "assistant",
17       "content": "To visualize the stock price of Netflix over the last five years, we will create a line chart that displays the adjusted closing price over time. This will help us understand the trend of Netflix's stock price in recent years.
18
19  <D3>
20  const endDate = new Date();
21  const startDate = new Date(endDate.getFullYear() - 5, endDate.getMonth(), endDate.getDate());
22  const filteredData = data.filter(d => d.Date >= startDate && d.Date <= endDate);
23  // .....
24  // [we skip the code for better readability]
25  // .....
26  svg.append('text')
27      .attr('x', vw / 2)
28      .attr('y', margin.top / 2)
29      .attr('text-anchor', 'middle')
30      .style('font-size', '16px')
31      .style('font-weight', 'bold')
32      .text('Netflix Stock Price (Last 5 Years)');
33
34  [xScale, yScale]
35  </D3>
36  This line chart shows the adjusted closing price of Netflix's stock over the last five years. The x-axis represents the time, while the y-axis represents the stock price. The line indicates the trend of the stock price, allowing you to observe any significant increases or decreases over this period.",
37     },
38     {
39       "role": "user",
40       "content": "
41  [Actions]
42  Action 1. Draw an image to present an intent, which is a steep and consistent upward trend in the stock price over a short period.
43  [Instruction]
44  Highlight all the time periods similar to this trend.",
45       "actions": [
46           {
47               "index": 1,
48               "type": "draw",
49               "element": "draft",
50               "data": "a steep and consistent upward trend in the stock price over a short period.",
51               "annotation": {
52                   "type": "draw",
53                   "paths": [
54                       // [we skipped the coordinates for readability]
55                       [[123, 540], [124, 536], [125, 533], ...]
56                   ]
57               }
58           }
```

```
59        ],
60        "instruction": {
61            "text": "Highlight all the time periods similar to this trend.",
62            "references": [
63                {
64                    "text": "this trend",
65                    "start": 42,
66                    "end": 52,
67                    "actionIdx": [
68                        1
69                    ]
70                }
71            ]
72        }
73    },
74    {
75        "role": "assistant",
76        "content": "To highlight periods of steep and consistent upward trends in Netflix's stock price, we will identify segments in the data where the adjusted closing price shows a significant increase over a short period. We will then visualize these segments on a line chart, using a different color to highlight these trends.
77
78  <D3>
79  function identifyUpwardTrends(data, threshold = 0.05, period = 5) {
80      const trends = [];
81      for (let i = 0; i < data.length - period; i++) {
82          const startPrice = data[i].AdjClose;
83          const endPrice = data[i + period].AdjClose;
84          const change = (endPrice - startPrice) / startPrice;
85          if (change > threshold) {
86              trends.push(data.slice(i, i + period + 1));
87          }
88      }
89      return trends;
90  }
91  // .....
92  // [we skip the code for better readability]
93  // .....
94  return [xScale, yScale]
95  </D3>
96
97  This visualization highlights periods of steep and consistent upward trends in Netflix's stock price over the last five years. The entire stock price trend is shown in light gray, while the identified upward trends are highlighted in red. This allows you to easily spot periods where the stock price increased significantly over a short period.",
98    }
99  ]
```

# II. Selected Chat Records with InterChat

Here, we additionally select a few chat records from our user study and an expert interview to elaborate on the multimodal interaction features and the system's capability to finish various visual analytics tasks.

For each record, we provide the name of the dataset (D.), the initial state of the visualization (Pre State), users' direct manipulation (Manipulation), the user prompt (Prompt), the inferred manipulation descriptors (Manipulation Descriptor), and the updated visualization (Post State). Our system will generate manipulation descriptors when users use the Free Drawing Tool. For Click, Box, and Lasso selection, the system will directly provide the selection data in JSON format. The highlighted text in the user prompts indicates the connected trigger phrases for the direct manipulations. When users draw on the canvas, their sketches appear as black lines (as shown in the figures below). However, when these images are fed into the language model, we convert the sketches to purple lines to better distinguish them from the visualization content.

| # | D. | Pre State | Direct Manipulation | Prompt | Manipulation Descriptor | Post State |
|---|---|---|---|---|---|---|
| 1 | Number of Death | 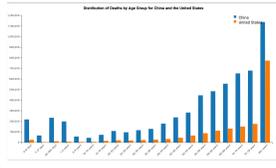 | 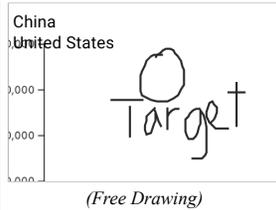<br>*(Free Drawing)* | Change the location of the legend to <u>this target</u>. | located near the top left of the chart, close to the 1,000,000 mark on the vertical axis, with a circle and the word "target" written in purple. | 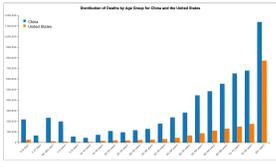 |
| 2 | Number of Death | 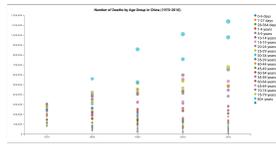 | 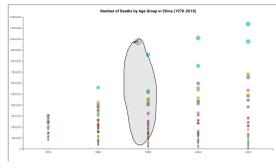<br>*(Lasso)* | Show me the gender difference for <u>the selected data using</u> a bar chart | (Lasso selection data) | 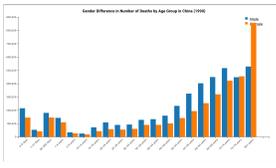 |
| 3 | Number of Death | 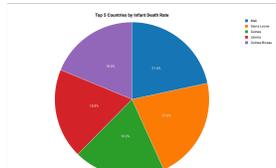 | 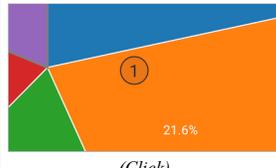<br>*(Click)* | Show me the gender distribution for <u>this country</u>. | (Click selection data) | 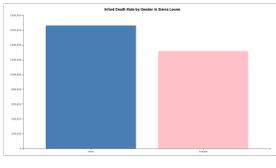 |
| 4 | Number of Death | 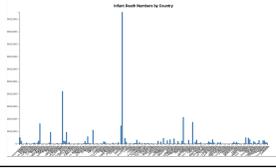 | 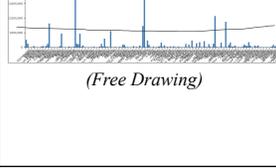<br>*(Free Drawing)* | Only keep the countries higher than <u>this value</u> and sort them from high to low. | a horizontal line between 1,000,000 and 2,000,000, approximately 1,200,000. | 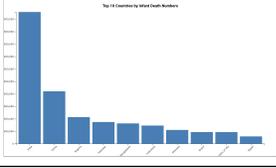 |
| 5 | Number of Death | 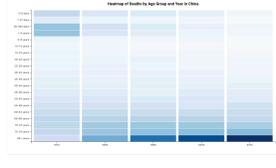 | 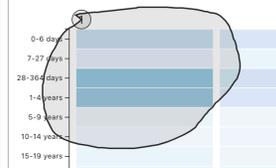<br>*(Lasso Selection)* | Compare the gender differences for <u>this data</u>. | (Lasso selection data) | 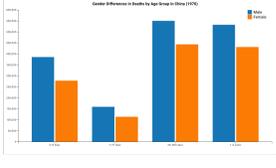 |
| 6 | Netflix Stock | 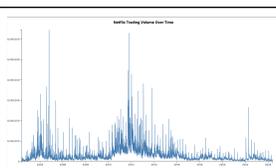 | 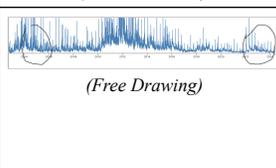<br>*(Free Drawing)* | Create chart for <u>these two time periods</u> and compare then side by side. | 2004 - 2006 and 2021 - 2023, marked by lines on the x axis of the chart. | 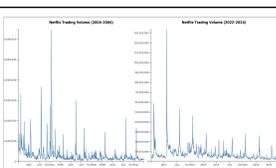 |
| 7 | Netflix Stock | 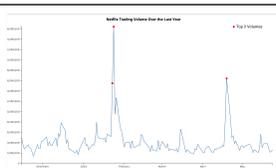 | 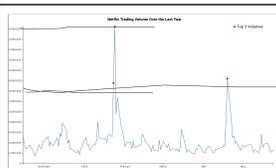<br>*(Free Drawing)* | Draw rules for <u>these lines</u> that I drew and annotate the values. | horizontal reference lines drawn to highlight the top three trading volumes on the chart, corresponding to the red points, and they annotate these specific volume levels. | 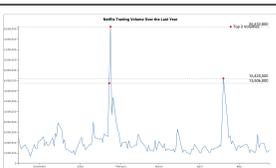 |

| # | Dataset | | | Instruction | Selection Data | Result |
|---|---------|---|---|-------------|----------------|--------|
| 8 | Netflix Stock | 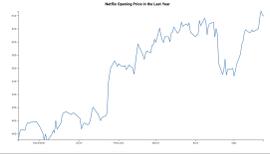 | 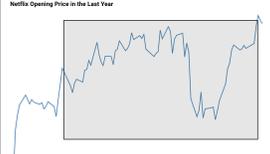<br>*(Box Selection)* | Do a bar chart that shows the trading volume over time in <u>the selected period</u>. | (Box selection data: x-axis range from late February to early May) | 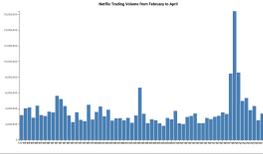 |
| 9 | Netflix Stock | 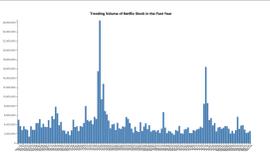 | 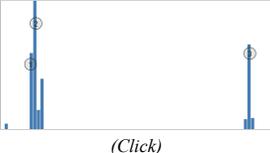<br>*(Click)* | Compare these <u>three days</u>' opening and closing stock prices | (Click selection data) | 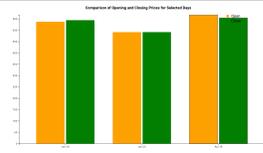 |
| 10 | Cars | 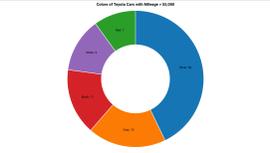 | 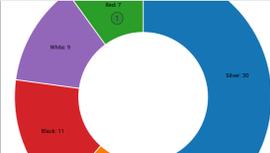<br>*(Click)* | What are the top 10 most expensive cars with <u>these two colors</u>? Please rank them using a horizontal bar chart. | (Click selection data) | 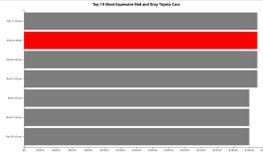 |
| 11 | Cars | 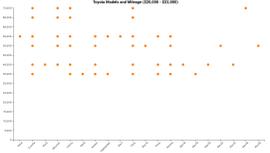 | 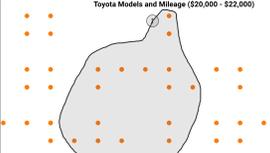<br>*(Lasso)* | What are the top 10 most expensive cars with <u>these two colors</u>? Please rank them using a horizontal bar chart. | (Lasso selection data) | 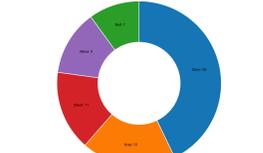 |
| 12 | uci-secom | 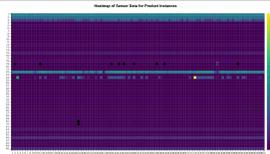 | 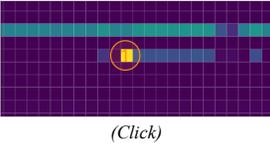<br>*(Click)* | Show me detailed information for <u>this sensor</u>. | (Click selection data) | 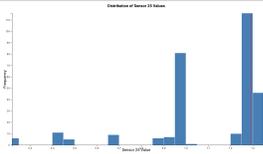 |
| 13 | Number of Death | | *((Open-Ended Initialization))* | Please show me something interesting about the data. | | *(To provide an interesting insight into the dataset, let's visualize the trend of death rates over the years for different age groups. This will help us understand how mortality rates have changed over time for various age categories…)*<br>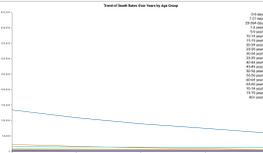 |
| 14 | Cars | | *((Open-Ended Initialization))* | Provide me with an overview of the dataset. | | *(To provide an overview of the dataset, I will create a bar chart that shows the count of cars by brand. This will give us a sense of the distribution of cars across different brands in the dataset….)*<br>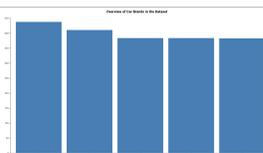 |

# III. Task Assignment for the User Study

In the user study, each participant was assigned 10 tasks to complete using the InterChat system. To minimize bias, we applied randomized task orders and task conditions, where in five of the tasks, users could only use the language input, and in the other five tasks, users could use both the language input and multimodal interaction features. Here is an example task document for the user study:

### InterChat: Tasks for Participant 1

Thank you for taking the time to participate in our user study! InterChat is a visual analytics system that incorporates multimodal interactions to better interpret your natural language intents to the LLMs. In this study, we kindly ask you to complete a series of tasks using our system, either **with** or **without** multimodal interactions.

#### Task 1

Please use the **stock** dataset to complete the following task (stock-4).

*Compare the trading volume values of two time periods that interest you. You can use stock price as a reference.*

Multimodal: You can use both the **language** input and **multimodel** interaction features to communicate with the system.

#### Task 2

Please use the **stock** dataset to complete the following task (stock-5).

*Identify the three days with the highest trading volume over the past year and compare these 3 days' opening and closing stock prices.*

Multimodal: You can use both the **language** input and **multimodel** interaction features to communicate with the system.

#### Task 3

Please use the **number of death** dataset to complete the following task (nd-5).

*Use a scatter plot to explore the number of deaths for different age groups in different years in China. Then, dive into a portion of the data to compare their gender differences.*

Multimodal: You can use both the **language** input and **multimodel** interaction features to communicate with the system.

#### Task 4

Please use the **number of death** dataset to complete the following task (nd-3).

*Identify the countries with the largest gender disparity in death rates.*

Multimodal: You can use both the **language** input and **multimodel** interaction features to communicate with the system.

#### Task 5

Please use the **stock** dataset to complete the following task (stock-2).

*Identify all the significant upward trends over the past five years, focusing on the periods with the most substantial increases.*

Language only: You should only use the **language** input to communicate with the system.

#### Task 6

Please use the **number of death** dataset to complete the following task (nd-1).

*Explore the number of infant deaths in France and Germany over the years.*

Language only: You should only use the **language** input to communicate with the system.

#### Task 7

Please use the **stock** dataset to complete the following task (stock-1).

*Find the monthly average data for the trading volume over the past three years.*

Language only: You should only use the **language** input to communicate with the system.

### Task 8

Please use the **stock** dataset to complete the following task (stock-3).

*Find a period of time in the last year with lots of fluctuations in stock price and explore the trading volume within this period.*

Multimodal: You can use both the **language** input and **multimodel** interaction features to communicate with the system.

### Task 9

Please use the **number of death** dataset to complete the following task (nd-4).

*Find out what countries have higher infant death rates, and filter the results to avoid visual clutter.*

Language only: You should only use the **language** input to communicate with the system.

### Task 10

Please use the **number of death** dataset to complete the following task (nd-2).

*Find the distribution of deaths by age group in China and the United States. You may need to move the legend to an appropriate location.*

Language only: You should only use the **language** input to communicate with the system.

Here is the task assignment matrix for 10 participants.

| Participant | Task 1 | Task 2 | Task 3 | Task 4 | Task 5 | Task 6 | Task 7 | Task 8 | Task 9 | Task 10 |
|---|---|---|---|---|---|---|---|---|---|---|
| 1 | stock-4, m | stock-5, m | nd-5, m | nd-3, m | stock-2, c | nd-1, c | stock-1, c | stock-3, m | nd-4, c | nd-2, c |
| 2 | stock-3, c | stock-2, c | stock-1, m | nd-1, c | nd-5, c | stock-4, m | nd-3, c | stock-5, m | nd-4, m | nd-2, m |
| 3 | nd-4, c | stock-2, m | nd-3, c | nd-5, m | nd-2, m | stock-1, m | stock-4, m | stock-5, c | stock-3, c | nd-1, c |
| 4 | stock-1, c | nd-2, c | stock-5, m | stock-4, m | nd-1, c | nd-3, c | stock-2, m | nd-4, m | stock-3, c | nd-5, m |
| 5 | nd-1, m | stock-5, m | nd-2, c | nd-5, m | nd-4, c | stock-1, m | nd-3, c | stock-4, c | stock-2, m | stock-3, c |
| 6 | stock-1, m | stock-5, c | stock-3, c | stock-2, c | nd-2, c | nd-3, m | nd-4, m | nd-1, m | nd-5, c | stock-4, m |
| 7 | stock-4, c | nd-2, c | stock-3, m | nd-3, m | stock-1, m | stock-2, m | nd-5, c | stock-5, c | nd-4, c | nd-1, m |
| 8 | nd-3, c | stock-2, m | stock-1, c | stock-5, m | nd-1, m | nd-2, c | nd-5, c | stock-4, c | nd-4, m | stock-3, m |
| 9 | nd-3, m | stock-3, m | stock-1, m | stock-4, c | nd-1, m | nd-2, m | stock-5, c | nd-4, c | stock-2, c | nd-5, c |
| 10 | nd-4, m | stock-2, c | stock-3, c | nd-3, m | stock-1, c | nd-1, c | nd-5, m | stock-4, m | nd-2, m | stock-5, c |

Here, each cell first indicates the task id, specified by the dataset name (`stock` or `nd`) and the task number (`1` to `5`). The suffix `c` or `m` in each cell indicates the task condition: `c` represents using the language input only, and `m` represents using both the language input and multimodal interactions.

*A. Scoring Results*

| ID | Participant | Task | Multimodal | Time (s) | # Conversation | # Retries | Intent Interpretation | Correct Answer |
|---|---|---|---|---|---|---|---|---|
| 1 | P1 | stock-1 | No | 40 | 1 | 0 | 2 (Correct) | 2 (Correct) |
| 2 | P1 | stock-2 | No | 242 | 3 | 3 | 1 (Partially Correct) | 0 (Incorrect) |
| 3 | P1 | nd-1 | No | 41 | 1 | 0 | 2 (Correct) | 2 (Correct) |
| 4 | P1 | nd-2 | No | 63 | 2 | 0 | 2 (Correct) | 2 (Correct) |
| 5 | P1 | nd-4 | No | 42 | 1 | 0 | 2 (Correct) | 2 (Correct) |
| 6 | P1 | stock-3 | Yes | 94 | 1 | 0 | 1 (Partially Correct) | 1 (Partially Correct) |
| 7 | P1 | stock-4 | Yes | 243 | 3 | 1 | 1 (Partially Correct) | 2 (Correct) |
| 8 | P1 | stock-5 | Yes | 122 | 2 | 1 | 2 (Correct) | 2 (Correct) |
| 9 | P1 | nd-3 | Yes | 304 | 4 | 0 | 2 (Correct) | 2 (Correct) |
| 10 | P1 | nd-5 | Yes | 143 | 2 | 0 | 2 (Correct) | 2 (Correct) |
| 11 | P2 | stock-2 | No | 360 | 4 | 1 | 2 (Correct) | 1 (Partially Correct) |
| 12 | P2 | stock-3 | No | 234 | 2 | 0 | 2 (Correct) | 2 (Correct) |
| 13 | P2 | nd-1 | No | 31 | 1 | 0 | 2 (Correct) | 2 (Correct) |
| 14 | P2 | nd-3 | No | 560 | 6 | 3 | 1 (Partially Correct) | 0 (Incorrect) |
| 15 | P2 | nd-5 | No | 304 | 2 | 0 | 2 (Correct) | 2 (Correct) |
| 16 | P2 | stock-1 | Yes | 31 | 1 | 0 | 2 (Correct) | 2 (Correct) |
| 17 | P2 | stock-4 | Yes | 70 | 2 | 1 | 2 (Correct) | 2 (Correct) |
| 18 | P2 | stock-5 | Yes | 72 | 3 | 1 | 2 (Correct) | 2 (Correct) |
| 19 | P2 | nd-2 | Yes | 181 | 4 | 3 | 1 (Partially Correct) | 1 (Partially Correct) |
| 20 | P2 | nd-4 | Yes | 110 | 2 | 0 | 2 (Correct) | 2 (Correct) |
| 21 | P3 | nd-4 | No | 22 | 1 | 0 | 2 (Correct) | 2 (Correct) |
| 22 | P3 | stock-2 | Yes | 91 | 2 | 0 | 2 (Correct) | 2 (Correct) |
| 23 | P3 | nd-3 | No | 203 | 3 | 0 | 2 (Correct) | 2 (Correct) |
| 24 | P3 | nd-5 | Yes | 122 | 2 | 0 | 2 (Correct) | 2 (Correct) |
| 25 | P3 | nd-2 | Yes | 124 | 3 | 1 | 2 (Correct) | 1 (Partially Correct) |
| 26 | P3 | stock-1 | Yes | 63 | 2 | 0 | 2 (Correct) | 2 (Correct) |
| 27 | P3 | stock-4 | Yes | 182 | 3 | 0 | 2 (Correct) | 2 (Correct) |
| 28 | P3 | stock-5 | No | 104 | 2 | 1 | 2 (Correct) | 2 (Correct) |
| 29 | P3 | stock-3 | No | 133 | 3 | 1 | 2 (Correct) | 2 (Correct) |
| 30 | P3 | nd-1 | No | 60 | 2 | 0 | 2 (Correct) | 2 (Correct) |
| 31 | P4 | stock-1 | No | 124 | 2 | 0 | 2 (Correct) | 1 (Partially Correct) |
| 32 | P4 | nd-2 | No | 243 | 3 | 1 | 1 (Partially Correct) | 2 (Correct) |
| 33 | P4 | stock-5 | Yes | 280 | 3 | 0 | 2 (Correct) | 2 (Correct) |
| 34 | P4 | stock-4 | Yes | 244 | 4 | 0 | 2 (Correct) | 2 (Correct) |
| 35 | P4 | nd-1 | No | 121 | 2 | 1 | 2 (Correct) | 1 (Partially Correct) |
| 36 | P4 | nd-3 | No | 450 | 2 | 1 | 2 (Correct) | 2 (Correct) |
| 37 | P4 | stock-2 | Yes | 182 | 2 | 2 | 1 (Partially Correct) | 1 (Partially Correct) |
| 38 | P4 | nd-4 | Yes | 61 | 2 | 0 | 2 (Correct) | 2 (Correct) |
| 39 | P4 | stock-3 | No | 180 | 2 | 0 | 2 (Correct) | 2 (Correct) |
| 40 | P4 | nd-5 | Yes | 482 | 2 | 0 | 2 (Correct) | 2 (Correct) |
| 41 | P6 | stock-1 | Yes | 61 | 2 | 0 | 2 (Correct) | 2 (Correct) |
| 42 | P6 | stock-5 | No | 123 | 2 | 0 | 2 (Correct) | 2 (Correct) |
| 43 | P6 | stock-3 | No | 242 | 3 | 2 | 1 (Partially Correct) | 2 (Correct) |
| 44 | P6 | stock-2 | No | 121 | 2 | 0 | 2 (Correct) | 2 (Correct) |
| 45 | P6 | nd-2 | No | 103 | 1 | 0 | 2 (Correct) | 2 (Correct) |
| 46 | P6 | nd-3 | Yes | 362 | 2 | 2 | 2 (Correct) | 0 (Incorrect) |
| 47 | P6 | nd-4 | Yes | 94 | 2 | 0 | 2 (Correct) | 2 (Correct) |
| 48 | P6 | nd-1 | Yes | 63 | 2 | 0 | 1 (Partially Correct) | 2 (Correct) |
| 49 | P6 | nd-5 | No | 400 | 3 | 2 | 1 (Partially Correct) | 2 (Correct) |
| 50 | P6 | stock-4 | Yes | 124 | 2 | 0 | 2 (Correct) | 2 (Correct) |
| 51 | P7 | stock-4 | No | 243 | 3 | 0 | 1 (Partially Correct) | 2 (Correct) |
| 52 | P7 | nd-2 | No | 300 | 5 | 0 | 2 (Correct) | 2 (Correct) |

| | | | | | | | | |
|---|---|---|---|---|---|---|---|---|
| 53 | P7 | stock-3 | Yes | 94 | 2 | 0 | 2 (Correct) | 2 (Correct) |
| 54 | P7 | nd-3 | Yes | 401 | 6 | 1 | 2 (Correct) | 2 (Correct) |
| 55 | P7 | stock-1 | Yes | 30 | 2 | 0 | 2 (Correct) | 2 (Correct) |
| 56 | P7 | stock-2 | Yes | 274 | 4 | 2 | 1 (Partially Correct) | 2 (Correct) |
| 57 | P7 | nd-5 | No | 151 | 3 | 1 | 1 (Partially Correct) | 2 (Correct) |
| 58 | P7 | stock-5 | No | 120 | 2 | 1 | 2 (Correct) | 2 (Correct) |
| 59 | P7 | nd-4 | No | 22 | 1 | 0 | 2 (Correct) | 2 (Correct) |
| 60 | P7 | nd-1 | Yes | 41 | 1 | 0 | 2 (Correct) | 2 (Correct) |
| 61 | P5 | nd-1 | Yes | 123 | 2 | 1 | 1 (Partially Correct) | 2 (Correct) |
| 62 | P5 | stock-5 | Yes | 102 | 3 | 0 | 2 (Correct) | 2 (Correct) |
| 63 | P5 | nd-2 | No | 241 | 4 | 2 | 1 (Partially Correct) | 1 (Partially Correct) |
| 64 | P5 | nd-5 | Yes | 153 | 2 | 0 | 2 (Correct) | 2 (Correct) |
| 65 | P5 | nd-4 | No | 242 | 2 | 0 | 2 (Correct) | 2 (Correct) |
| 66 | P5 | stock-1 | Yes | 44 | 1 | 0 | 2 (Correct) | 2 (Correct) |
| 67 | P5 | nd-3 | No | 483 | 3 | 3 | 1 (Partially Correct) | 0 (Incorrect) |
| 68 | P5 | stock-4 | No | 122 | 2 | 0 | 2 (Correct) | 2 (Correct) |
| 69 | P5 | stock-2 | Yes | 104 | 2 | 0 | 1 (Partially Correct) | 2 (Correct) |
| 70 | P5 | stock-3 | No | 123 | 2 | 0 | 2 (Correct) | 2 (Correct) |
| 71 | P8 | stock-2 | Yes | 180 | 3 | 0 | 2 (Correct) | 2 (Correct) |
| 72 | P8 | stock-1 | No | 64 | 2 | 0 | 1 (Partially Correct) | 2 (Correct) |
| 73 | P8 | stock-5 | Yes | 181 | 3 | 1 | 1 (Partially Correct) | 2 (Correct) |
| 74 | P8 | nd-3 | No | 300 | 4 | 3 | 1 (Partially Correct) | 2 (Correct) |
| 75 | P8 | nd-1 | Yes | 94 | 2 | 1 | 2 (Correct) | 2 (Correct) |
| 76 | P8 | nd-2 | No | 91 | 2 | 0 | 2 (Correct) | 2 (Correct) |
| 77 | P8 | nd-5 | Yes | 340 | 4 | 1 | 1 (Partially Correct) | 2 (Correct) |
| 78 | P8 | stock-4 | Yes | 172 | 2 | 0 | 2 (Correct) | 2 (Correct) |
| 79 | P8 | nd-4 | Yes | 61 | 2 | 0 | 2 (Correct) | 2 (Correct) |
| 80 | P8 | stock-3 | No | 90 | 2 | 0 | 2 (Correct) | 2 (Correct) |
| 81 | P9 | stock-3 | Yes | 382 | 3 | 1 | 1 (Partially Correct) | 1 (Partially Correct) |
| 82 | P9 | stock-1 | Yes | 181 | 1 | 2 | 1 (Partially Correct) | 2 (Correct) |
| 83 | P9 | stock-4 | No | 243 | 2 | 0 | 2 (Correct) | 2 (Correct) |
| 84 | P9 | nd-1 | No | 202 | 2 | 0 | 2 (Correct) | 1 (Partially Correct) |
| 85 | P9 | nd-2 | Yes | 194 | 1 | 1 | 2 (Correct) | 2 (Correct) |
| 86 | P9 | stock-5 | No | 53 | 2 | 0 | 2 (Correct) | 2 (Correct) |
| 87 | P9 | nd-4 | No | 182 | 2 | 2 | 1 (Partially Correct) | 2 (Correct) |
| 88 | P9 | stock-2 | No | 214 | 5 | 2 | 1 (Partially Correct) | 1 (Partially Correct) |
| 89 | P9 | nd-5 | No | 243 | 3 | 0 | 2 (Correct) | 2 (Correct) |
| 90 | P9 | nd-3 | Yes | 60 | 1 | 0 | 2 (Correct) | 2 (Correct) |
| 91 | P10 | nd-4 | Yes | 124 | 2 | 0 | 2 (Correct) | 2 (Correct) |
| 92 | P10 | stock-2 | No | 201 | 3 | 3 | 2 (Correct) | 1 (Partially Correct) |
| 93 | P10 | stock-3 | No | 240 | 4 | 2 | 2 (Correct) | 2 (Correct) |
| 94 | P10 | stock-1 | No | 74 | 2 | 0 | 2 (Correct) | 2 (Correct) |
| 95 | P10 | nd-1 | No | 21 | 1 | 0 | 2 (Correct) | 2 (Correct) |
| 96 | P10 | nd-5 | Yes | 120 | 2 | 0 | 2 (Correct) | 2 (Correct) |
| 97 | P10 | stock-4 | Yes | 62 | 2 | 0 | 2 (Correct) | 2 (Correct) |
| 98 | P10 | nd-2 | Yes | 151 | 2 | 0 | 1 (Partially Correct) | 2 (Correct) |
| 99 | P10 | stock-5 | No | 60 | 1 | 0 | 2 (Correct) | 2 (Correct) |
| 100 | P10 | nd-3 | Yes | 172 | 2 | 0 | 2 (Correct) | 2 (Correct) |



\end{document}